\documentclass{aastex631}
\usepackage{graphicx} 
\usepackage{amsmath} 
\usepackage{longtable}

\shorttitle{}

\begin{document}

\title{The Metal Content of Resolved Galaxies}

\author[0000-0003-0736-7609]{Lidia N. Makarova}
\affiliation{Special Astrophysical Observatory of the Russian Academy of Sciences, Nizhnij Arkhyz, Karachay-Cherkessia 369167, Russia: lidia.makarova@gmail.com}

\author[0009-0008-5135-071X]{Denis G. Purytin}
\affiliation{Special Astrophysical Observatory of the Russian Academy of Sciences, Nizhnij Arkhyz, Karachay-Cherkessia 369167, Russia}
\affiliation{Saint Petersburg State University, Universitetskaya Embankment 7/9, St. Petersburg 199034, Russia}

\author[0000-0002-9291-1981]{R. Brent Tully}
\affiliation{Institute for Astronomy, University of Hawaii, 2680 Woodlawn Drive, Honolulu, HI 96822, USA}

\author[0000-0002-5259-2314]{Gagandeep S.\ Anand}
\affiliation{Space Telescope Science Institute, 3700 San Martin Drive, Baltimore, MD 21218, USA}

\begin{abstract}
We present a homogeneous metallicity analysis of old stellar populations in Local Volume (LV)
galaxies using data from the CMDs/TRGB catalog of the Extragalactic Distance Database (EDD;
http://edd.ifa.hawaii.edu), which provides uniformly measured TRGB distances and PSF photometry
for resolved stars in over 500 nearby galaxies observed with the Hubble Space Telescope.
We apply the calibration of \citet{1993ApJ...417..553L} to estimate the mean metallicity
[Fe/H] from the $(V-I)$ color of the red giant branch (RGB) at $M_I = -3.5$ mag obtaining 
reliable measurements for 334 galaxies out of an initial set of 558. The RGB colors were 
derived by locating the maximum stellar density in the ($M_I, (V-I)$) diagram, smoothed with a 
Gaussian kernel and refined via Monte Carlo simulations (500–1000 realizations), yielding 
typical uncertainties of about 0.03 mag. Our results show that most galaxies lie within the
color range $(V–I) = 1.22–1.74$ of the original calibration, corresponding to metal-poor 
systems typical of dwarfs, with the overall metallicity distribution peaking at [Fe/H] = 
$-1.89\pm0.03$ dex. We find a pronounced luminosity–metallicity relation across a wide
magnitude range, from faint dwarfs ($M_B > -7$) mag to giant galaxies ($M_B < -18$), 
described by the regression $\text{[Fe/H]} = -2.6 - 0.075 \, M_B$. Both dwarf and giant 
galaxies follow the same relation, though ACS fields for giants often sample outer, more 
metal-poor regions. Morphologically, early-type dwarf spheroidals exhibit systematically higher
mean metallicities than late-type dwarf irregulars.
\end{abstract}


\section{Introduction}

Metal content ([Fe/H], Z) is a one of the key parameters
characterizing galaxy properties. It contains information about the processes affecting galaxy
formation and evolution. Various physical processes are
connected to the mean and variation of metallicity in a 
galaxy: primarily, star formation processes, their intensity 
and duration in different regions of the galaxy and different 
environments. The accretion of intergalactic gas and the 
galaxy's ability to retain enriched matter formed during 
stellar evolution can also have a significant impact.

There are a number of widely used methods for estimating the metal content of a galaxy. One of them is spectroscopy of
bright starforming H\,II regions (for example, 
\citet{1989ApJ...347..875S}, \citet{2006ApJ...637..269V}, \citet{2023MNRAS.518.5893K}, \citet{2024A&A...682L..11S}).
Oxygen abundance determined from an H\,II region can be 
translated into a metallicity value
and is directly related to the most recent, ongoing episode of
star formation.

Alternatively, spectroscopic measurements of the metallicity of 
individual red giant stars within galaxies provide estimates of 
the metal abundances of old stellar populations. These 
measurements have become possible for nearby galaxies resolved into stars thanks to advances in
observational capabilities (see, for example, the studies of 
\citet{2012ApJ...752...45T}, \citet{2013ApJ...779..102K}, \citet{2024ApJ...972..133C}).

Nearby galaxies can be resolved into
individual stars with Hubble Space Telescope (HST) imaging such that sufficiently deep color-
magnitude diagrams (CMD) can be obtained. From these CMD it is possible to
determine the dependence of the stellar population metallicities
on their age by fitting theoretical stellar isochrones to the
observational data (for example, \citet{2011ApJ...739....5W}, \citet{2021MNRAS.502.1623M}; see also the review of \citet{2022NatAs...6...48A}).

More generally, for nearby resolved galaxies there is a well-known and relatively easily applicable method for estimating the average metallicity of the upper part of the red giant branch near its tip using the RGB color index, closely related to the TRGB distance indicator (\citet{1993ApJ...417..553L}, \citet{2008MNRAS.384.1544S}).

One of the most well-established and pronounced correlations involving the fundamental properties of galaxies is the relation between galaxy mass and metallicity—the mass–metallicity relation (MZR). There is also a sparser, yet still evident, luminosity–metallicity relation (LZR). Both relations clearly demonstrate that more massive and more luminous galaxies possess higher metal content (see, for example, \citet{2013ApJ...779..102K}, \citet{2025OJAp....8E.142P}, \citet{2026MNRAS.545f2011K}).

In an early study, \citet{1986ApJ...303...39D} reported a tight correlation between metallicity and luminosity in dwarf galaxies. The authors found a clear trend spanning approximately eight magnitudes, $L \propto Z^z$, with z=2.5. The observed relation between stellar mass and metallicity provides an important basis for models of chemical enrichment in galaxies, as well as for understanding intergalactic gas inflow and the mechanisms governing the retention and outflow of enriched gas.

Despite numerous studies (see the works mentioned above) that estimate both the gas-phase oxygen abundance in galaxies of various types and at different distances, as well as the metallicity of their old stellar populations, many questions remain unresolved. Metallicity measurements are based on samples of varying and sometimes limited sizes. The trends and issues in the mass–metallicity relation, particularly regarding galaxies of different morphological types located at varying distances and, crucially, in different environments, are still not fully understood. Comparisons of stellar masses and metallicities derived using different methods also often lead to conflicting results.

In this paper, we determine, for the first time, the metallicity from the color index of the upper part of the red giant branch for the entire sample of nearby galaxies that are resolved into individual stars (mostly dwarf galaxies) using stellar photometry data from the Hubble Space Telescope available in the Extragalactic Distance Database (EDD: \texttt{http://edd.ifa.hawaii.edu/}; \citealt{2009AJ....138..323T, 2021AJ....162...80A}).

\section{Red Giant Branch color-metallicity relation}

\citet{1990AJ....100..162D} obtained V, I CCD-photometry of red giants in eight Galactic globular clusters that cover a wide range of metal abundance, which allowed them to construct $(M_I, (\text{V} - \text{I}))$
color-magnitude diagrams and study abundance variations.
The greatest sensitivity to variations occurs at higher luminosities of the RGB.
Assuming metallicity estimates from the integrated spectroscopy of these calibrating globular clusters \citep{1985ApJ...293..424Z, 1988AJ.....96...92A}, the authors derived an empirical relation between metallicity [Fe/H] and the $(\text{V} - \text{I})_0$ color index of the RGB at the level of $M_I = -3.0^m$. This absolute magnitude level was selected as a compromise, as some sparse globular clusters may have red giant branches that are not populated at the highest luminosities. This calibrating relation was obtained for the metallicity range [Fe/H] from $-0.71$ dex (47 Tuc) to $-2.17$ dex (M15).

Later, \citet{1993ApJ...417..553L} calibrated this relation for the level of $M_I = -3.5^m$. The authors encountered the problem of significantly larger errors for the red giant stars in galaxies than for the globular clusters. Therefore, the previous color-metallicity relation was recalculated using the data on the calibrating Galactic globular clusters from \citet{1990AJ....100..162D} for the absolute magnitude level $M_I = -3.5^m$ (i.e., $0.4 - 0.6^m$ below the TRGB), valid within a similar metallicity range. This calibration allowed the method to be applied to shallower color-magnitude diagrams where the photometric limit may be located closer to the tip of the red giant branch. The resulting relation is following:

\begin{equation}
    \text{[Fe/H]} =-12.64 + 12.6 \cdot (\text{V}-\text{I})_{M_I=-3.5} - 3.3 \cdot (\text{V}-\text{I})_{M_I=-3.5}^2
    \label{formula:main-equation}
\end{equation}
This relation was calculated for the range $–2.17 \le \text{[Fe/H]} \le –0.71$ dex, which
corresponds to $1.22 \leq (\text{V}-\text{I}) \leq 1.74~^{m}$. The formulation is widely 
used to estimate an average value of
the metallicity of individually resolved red giant stars in nearby galaxies. We have used the relation (\ref{formula:main-equation}) to estimate the metallicity of our large sample of Local Volume\footnote{CMD  reach sufficiently deeper than the TRGB in single orbit observations with HST for galaxies within 10~Mpc, a domain we call the Local Volume.} galaxies observed with HST.

\section{Photometric data}
Hubble Space Telescope provides us with a unique opportunity to obtain images of nearby galaxies that are resolved to individual stars. Since the telescope's launch, several hundred Local Volume galaxies, mostly dwarfs, have been imaged with WFPC2 and then with ACS and WFC3 cameras. The images obtained in at least two broadband filters (in particular, in F606W or F555W and F814W) allow us to use the method of \citet{1993ApJ...417..553L} to determine metallicity.
A number of HST programs related to TRGB distance determinations provide us with a deep homogeneous set of images (see, for example, HST proposals 15922, 13442, 12546, 10235, 10210, 9771, 8601, 8192). 
The publicly accessible EDD contains the catalog CMDs/TRGB which provides us with uniformly measured high-precision TRGB distances on a common scale and with tables of PSF-photometry of resolved stars for the vast majority of galaxies observed with HST in the Local Volume \citep{2021AJ....162...80A}. This catalog has 558 entries for now.

Most of the CMDs/TRGB have been obtained with the ACS/WFC. Early observations made with WFPC2 represent a prominent part of the data on the nearest galaxies. The color-magnitude diagram (CMD) is provided for resolved stars of each galaxy. Predominantly, the CMD have been constructed using F606W and F814W filters in the form (F606W-F814W), F814W, but F555W also may be used to measure the necessary color. The photometric table that is available for download for each galaxy in the catalog provided us with the F606W (or F555W) and F814W apparent magnitudes and photometric errors transformed into Johnson-Cousins V and I values necessary for the purpose of [Fe/H] metallicity estimation, as described in previous section (see also Section 2 of \citet{2021AJ....162...80A} for detailed data description).  Stellar PSF-photometry of resolved stars of every galaxy in the CMDs/TRGB catalog was performed with the DOLPHOT \citep{2000PASP..112.1397D, 2016ascl.soft08013D} package for crowded-field photometry for HST images (for detailed description, see Section 3 of \citet{2021AJ....162...80A}). We then selected the stars of a good photometric quality from the output of the DOLPHOT photometry (*.phot file), using type of stellar object, signal-to-noise level, sharpness and crowding parameters: $t \le 2, s/n_V \ge 4, s/n_I \ge 4, (sharp_V + sharp_I)^2 \le 0.075, flag_V \le 2 \text{*}, flag_I \le 2\text{*}, crowd_V + crowd_I \le 0.8\text{*}$. The parameters marked with [*] were missing from some of the tables produced using older versions of DOLPHOT. In such cases, these items were disregarded. 
The Galactic extinction was taken into account following \citet{2011ApJ...737..103S}. 

Some of the galaxies in the sample were found to be too far for a TRGB distance estimation, or the red bounding edge compromised the ability to measure the $(\text{V} - \text{I})$ color. Such galaxies were excluded from further analysis. In sum, 445 measurements of the red giant branch $(\text{V}–\text{I})$ color index were obtained, giving estimates of metallicity.
Out of these, 111 measurements were discarded due to insufficient photometric depth. Although the photometry was deep enough to detect the tip of the red giant branch (TRGB), at the level of $M_I = –3.5^m$ the photometric limit could significantly bias the results.
Among the remaining 334 galaxies, we categorized the data as follows:
\begin{itemize}
    \item 218 galaxies: Reliable measurements with no significant influence from the photometric limit.
    \item 38 galaxies: The photometric limit is relatively close and may have a minor impact on the result; however, this impact is considered negligible.
    \item 58 galaxies: The measured color index falls outside the valid range $(\text{V}–\text{I}) \in [1.22, 1.74]$ of the adopted calibration of \citet{1993ApJ...417..553L}, and these results provide only an approximate estimate.
    \item 20 galaxies: Both negative conditions are present -- the photometric limit is close, and the $(\text{V} - \text{I})$ color is outside the calibration range.
\end{itemize}

While the latter three categories cannot be used as reliable metallicity estimations, they still provide us with the preliminary approximations of [Fe/H] and are thus presented in the resulting table.

\section{Metallicity estimation technique}
\label{section:technique}

\subsection{The color estimation criteria}
According to equation (\ref{formula:main-equation}), to measure [Fe/H] for a galaxy in our sample we need an appropriate estimation of the $(\text{V}-\text{I})$ color of the red giant branch at the level $M_I = -3.5^m$. However, the measurement of a mean or median value of the color index $(\text{V}-\text{I})$ at a fixed luminosity level is not straight forward.
The values depend on the chosen $(\text{V}-\text{I})_{M_I=-3.5}$ range where the statistical counting of stars is performed. An appropriate range could be questionable in galaxies with different populations of different ages and metallicities.
Therefore, a different approach was applied. The color index $(\text{V}-\text{I})_{M_I=-3.5}$ was determined as a maximum of the stellar density distribution at the $(M_I, \text{V}-\text{I})$ diagram.
Even so, this approach turned out to be ineffective for a number of galaxies, where the number of red giants was insufficient for a reliable approximation of the density. In such situations, a parabola approximation of the RGB was applied.

\subsection{Constructing a density profile}
\label{section:denisty_profile}
To obtain the position of the maximum of the stellar density profile at the level of $M_I=-3.5^m$, the uncertainties associated with both the photometric errors and the distance modulus uncertainties should be taken into account. Therefore, the profile width was determined as the combined value of these errors.
To construct the profile, we used smoothing with a Gaussian kernel, described by the formula:

\begin{equation}
    \rho(x) = \sum_{i=1}^{N} \frac{1}{\sigma \sqrt{2 \pi}} \exp{\left(- \frac{(x - c_i)^2}{2 \sigma^2}\right)}
    \label{formula:kernel-dencity}
\end{equation}
where $\{c_i\}_{i \in 1}^N$ are the color indices of the profile stars, and the parameter $\sigma$ characterizes the degree of smoothing. The average photometric error of the color index of the profile stars was taken as 1~sigma.

\begin{figure}[p]
    \center
    \includegraphics[width=0.4\textwidth]{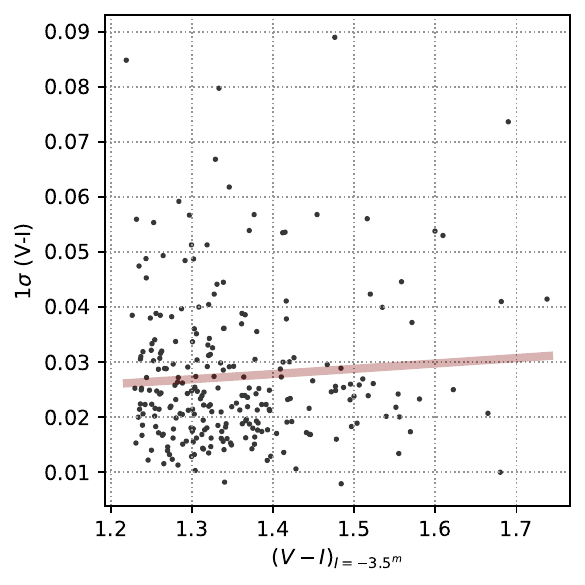}
    \includegraphics[width=0.4\textwidth]{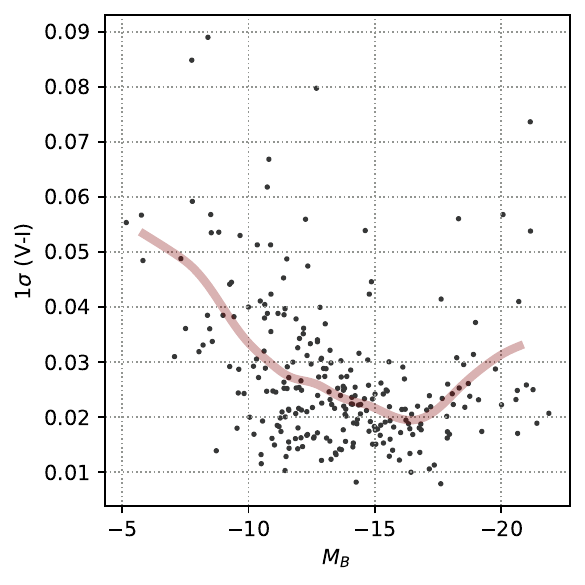}
    \caption{\textbf{Left}: Color index confidence interval vs. color index iteself. \textbf{Right}: Color index confidence interval vs. $M_B$. Bright galaxies have a wide range of metallicities, making the determination of a single value less certain, which explains some increase in the confidence interval at the right side of the graph. The light red lines illustrate the data smoothed with splines to enhance visual clarity.}
    \label{fig:error_distribution_vi}
\end{figure}

\begin{figure}[p]
    \center
    \includegraphics[width=0.4\textwidth]{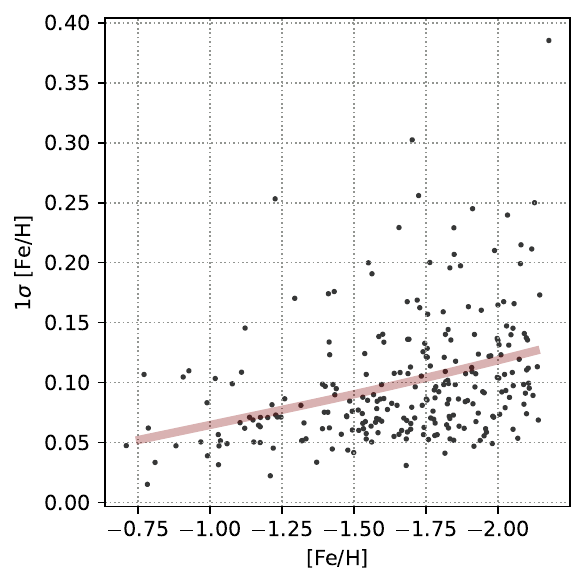}
    \includegraphics[width=0.4\textwidth]{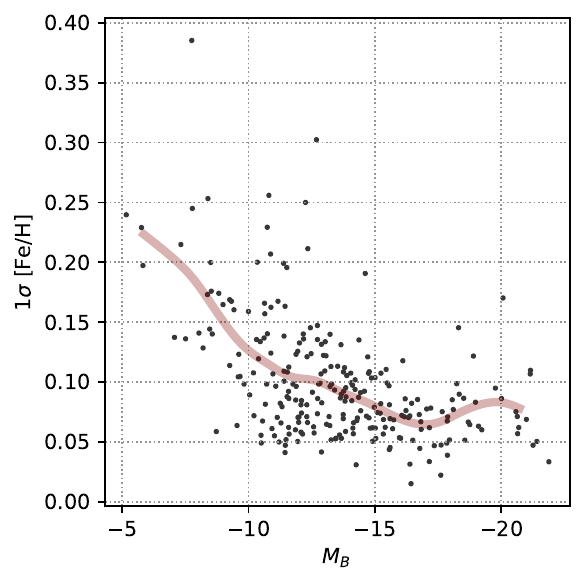}
    \caption{\textbf{Left}: Metallicity confidence interval vs. metallicity itself. High metallicities fall on the flatter part of the equation (\ref{formula:main-equation}), which explains the smaller confidence intervals on average. \textbf{Right}: Metallicity confidence interval vs. $M_B$. The light red lines illustrate the data smoothed with splines to enhance visual clarity.}
    \label{fig:error_distribution_feh}
\end{figure}

The position of the maximum of the resulting density profile was used as a preliminary estimation of the $(\text{V}-\text{I})_{M_I=-3.5}$ value. However, this estimate alone does not provide a robust color value, as the density profile may be sensitive to small variations in the stellar distribution near the maximum. Therefore, to obtain a reliable estimate, we applied a Monte Carlo method. In each of 500–1000 realizations, the magnitudes and color indices of the stars were randomly changed within their photometric uncertainties. For each realization, the position of the density maximum was determined. The final estimate of the RGB color was taken as the mean value of the maximum positions obtained across all realizations. The standard deviation of these values was adopted as the uncertainty of the result. Resulting color uncertainties vs. color is demonstrated at the left panel of the Fig.~\ref{fig:error_distribution_vi}. The mean 1$\sigma$ uncertainty is practically independent on the color value and on average is about 0.03. The color index confidence interval vs. $M_B$ is shown at the right panel of Fig.~\ref{fig:error_distribution_vi}. Most objects in our sample are typical dwarfs and have agreeably low color index confidence values, whereas faintest and brightest ``tails" have on average higher uncertainties. With brighter galaxies, the fields chosen for the determination of TRGB distances were frequently at halo locations, exploring a wider range of metallicities, making the determination of an average value less certain. A consequence is increased uncertainty in values at the right end of the plot. The study of \citet{2008MNRAS.384.1544S} shows a similar trend in the dispersion of RGB color measured at $M_I = -3.5$ versus absolute magnitude in the V band for dSphs (dots) and dIrrs in the Local Volume. Although their sample is small, the dependence is more pronounced there, as the dispersion is measured using artificial stars, so reflects solely the dispersion range coming from the age and metallicity mix of the galaxies at the different ranges of absolute magnitudes.
Fig.~\ref{fig:error_distribution_feh} demonstrates the metallicity confidence interval vs. metallicity itself. Most galaxies of the sample have 1$\sigma$ uncertainty in [Fe/H] less than 0.2. High metallicities fall on the flatter part of equation (\ref{formula:main-equation}), which explains on average the smaller confidence intervals in this part of the plot.

\begin{figure}[h]
    \center
    \includegraphics[width=0.49\textwidth]{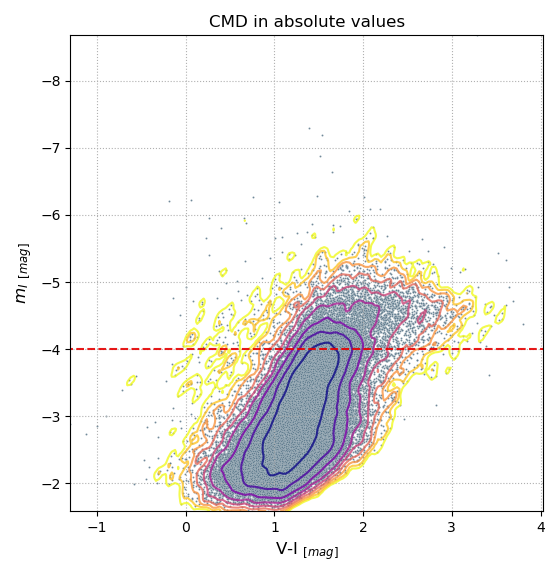}
    \includegraphics[width=0.49\textwidth]{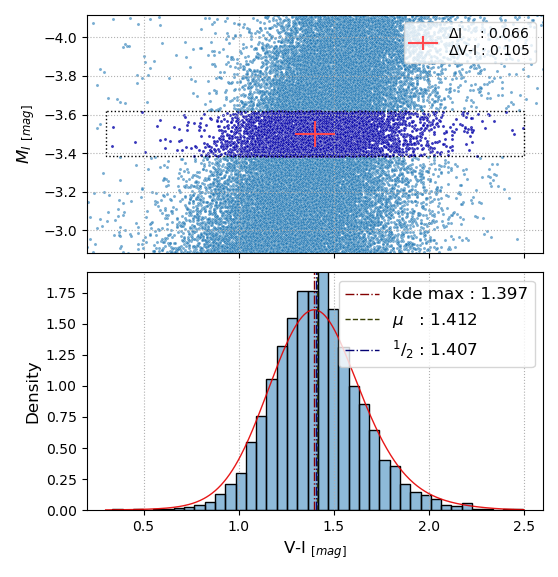}
    \caption{\textbf{Left}: Color–magnitude diagram for galaxy PGC 1034, as produced by our software. The x-axis shows the color index (V–I)? corrected for Galactic extinction, and the y-axis shows the absolute magnitude in the I-band. The horizontal red line marks the level of $M_I = -4^m$, approximately corresponding to the tip of the red giant branch (TRGB). Colored lines on the graph represent stellar isodensity contours on the color-magnitude diagram to improve visual clarity. \textbf{Right}: Density profile of the red giant branch at $M_I = -3.5^m$. The top panel shows a fragment of the RGB with a rectangular selection highlighting the stars used for constructing the profile. The bottom panel presents the histogram of the stellar distribution along the profile, along with the estimated density curve. The positions of the maximum, the mean and the median are indicated. }
    \label{fig:cmd_and_profile_examples}
\end{figure}

Fig.~\ref{fig:cmd_and_profile_examples} demonstrates an example of the color estimation for the case of IC\,10 (PGC\,1034). These diagrams were generated with our software, which required input of the photometric data together with distance and color extinction. In the case of IC\,10, the red giant branch is overlain by the asymptotic giant branch, making the TRGB “step” difficult to identify visually. Nevertheless, the branch itself is relatively clean, with little contamination on either side, so that the density maximum coincides closely with both the mean and the median color. This is not always the case for other galaxies. The right-hand panel illustrates the construction of the branch “profile”: stars are selected at $M_I = -3.5^m$ within the range defined by the photometric errors and distance uncertainty. From these data a density profile is built, and its maximum is used as the estimate of the branch color.

\subsection{Sparse samples}
\label{section:sparse_samples}

For galaxies with a sparse red giant branch, where the constructed density profile is poor, an alternative approach was taken. The RGB position was manually identified on the CMD, excluding stars that deviated significantly (usually from 1 to 7\% of the total number). The data were approximated by a parabola $(\text{V}–\text{I}) = \alpha I^2 + \beta I + \gamma$ using the least squares method. The confidence interval was estimated taking into account the spread of the approximation parameters and the uncertainty in the distance determination.

\subsection{Milky Way contamination}

\begin{figure}[h]
    \center
    \includegraphics[width=0.49\textwidth]{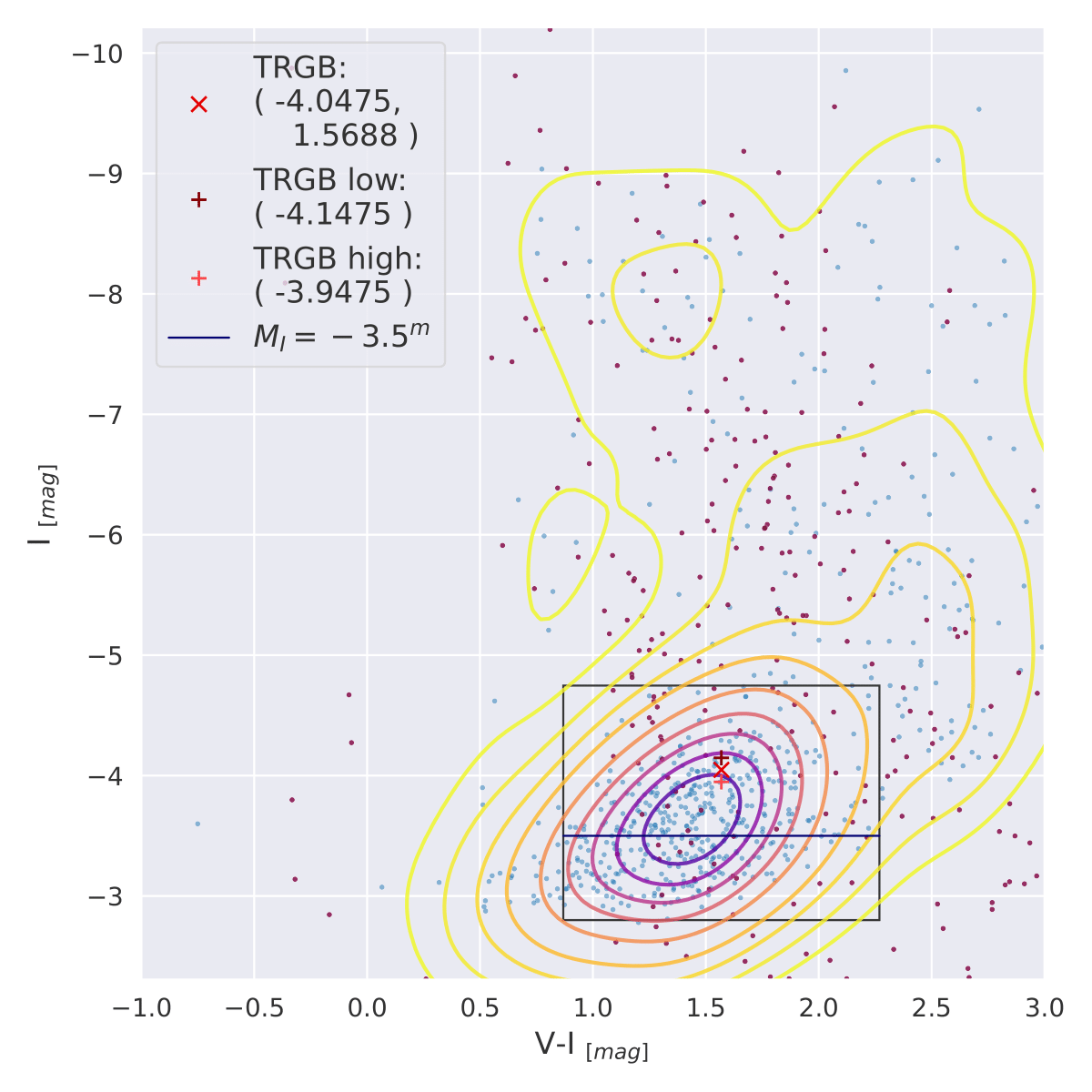}
    \includegraphics[width=0.49\textwidth]{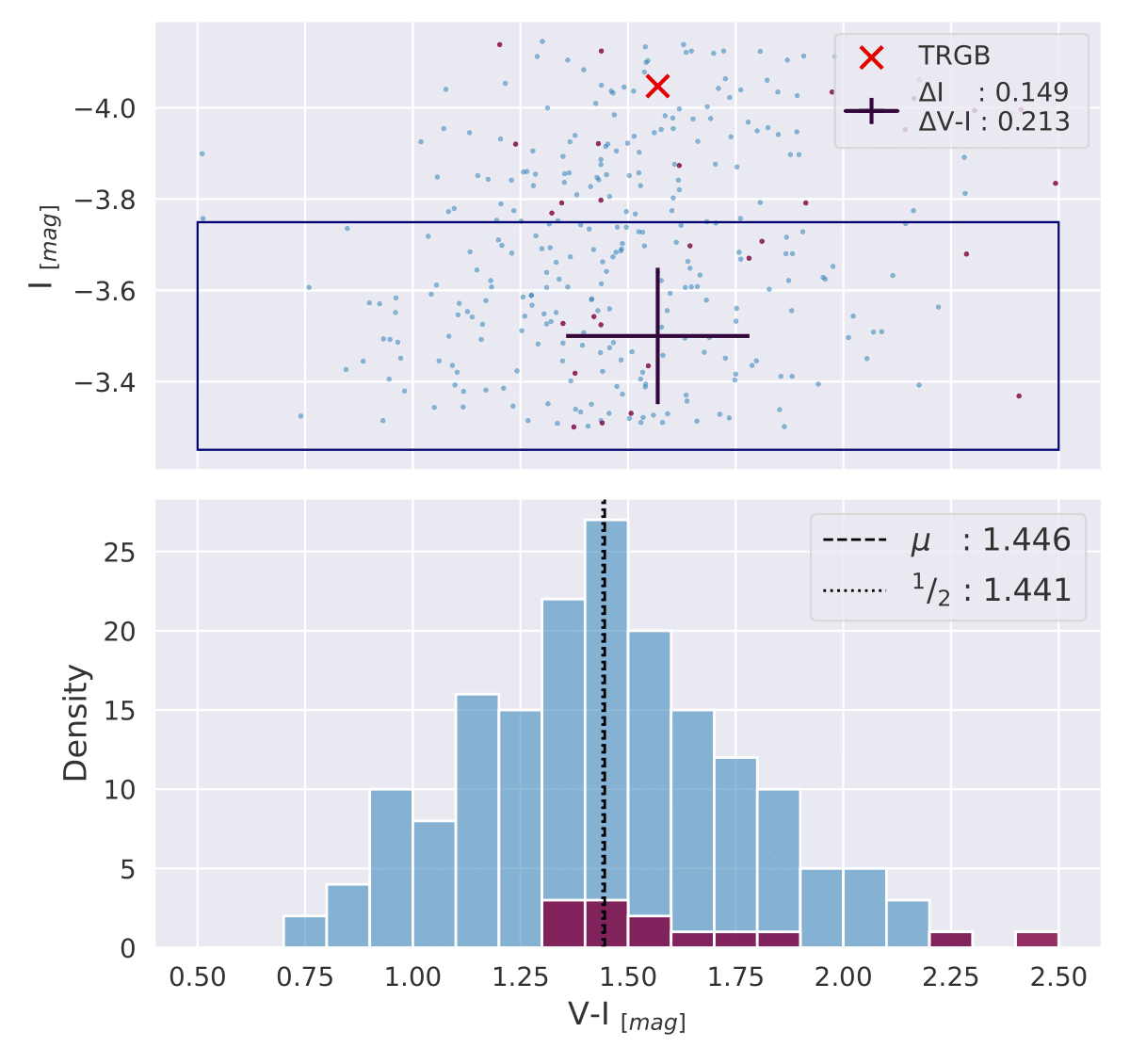}
    \caption{\textbf{Left}: Color–magnitude diagram showing stars from the target galaxy PGC166172 (blue points) together with foreground contamination from the Milky Way, and simulated Milky Way stars (red points) generated using the TRILEGAL software. Stellar density isocontours are overlaid. As can be seen, the model distribution does not match the observed foreground contamination in detail. \textbf{Right}: Red giant branch profile for the same galaxy. Red points again represent Milky Way stars. The bottom panel shows the corresponding histogram of the stellar distribution. The influence of Milky Way stars on the color index appears to be minor.}
    \label{fig:trilegal_screenshots}
\end{figure}

In some cases where the Milky Way contamination is serious it is difficult to clearly distinguish the target galaxy stars from the Milky Way foreground population. For these, we estimated the possible level of contamination in order to account for it. For each galaxy, we simulated the expected foreground stellar population of the Milky Way using the TRILEGAL software package \citep{2005A&A...436..895G} with respective Galactic coordinates, given the instrument that was used (see Fig.~\ref{fig:trilegal_screenshots} for example).
For the reasons of statistical accuracy, we chose not to subtract the TRILEGAL-simulated star distribution from the observed data. Though the TRILEGAL simulated CMD of the Milky Way looks reasonable in the most cases, the simulated distribution somewhat differs from the observed one, both in spatial arrangement within the CMD and in total star number.
Therefore, we analyzed such galaxies individually. In cases where the number of simulated Milky Way stars in the TRGB region was relatively high, we manually isolated the RGB branch and approximated it with a parabola, as described in Section \ref{section:sparse_samples}. If the contamination level was negligible, we applied the standard methodology outlined in Section \ref{section:denisty_profile}.

\section{The software tools}
The described procedures were implemented as a Python3 software package with a graphical user interface based on Qt5. The package is designed for the rapid and homogeneous processing of photometric data, utilizing Matplotlib for visualization. The program takes a standardized CSV file with a photometric catalog as input, along with user-provided values for color extinction and distance modulus.

A key feature is the interactive selection of the stellar sample. The user can define a region of interest both by its spatial coordinates and by the local stellar density, which allows for the effective exclusion of foreground stars or crowded galactic regions. The software then calculates the RGB color using the two methods described in this work, allowing the user to vary the analysis parameters and assess the stability of the solution. The final output includes a PDF report with summary plots and a machine-readable JSON file with the numerical results.
The program code is publicly available on \href{https://github.com/DenisPur/RGB-Color}{GitHub}
with the user manual included. A more detailed description of the package will be given in a forthcoming paper (Purytin D., Makarova L., 2026, in preparation).

\section{Metallicities of the galaxies in the Local Volume}

The results of our measurements of the metallicity [Fe/H] for the all possible Local Volume galaxies from the EDD database are given in the Table~\ref{tab:main}. The table contains the following columns: (1) -- PGC number; (2) -- galaxy name; (3, 4) -- equatorial coordinates (J2000); (5) -- TRGB distance in Mpc, taken from the EDD; (6) -- absolute stellar magnitude \citep{2013AJ....145..101K}; (7) -- morphological type \citep{2012AstBu..67..115K}; (8) -- extinction E(B-V) \citep{2011ApJ...737..103S}; (9) -- $(\text{V}-\text{I})_{M_I=-3.5}$ measured according to the procedure described in Section \ref{section:technique} (Galactic extinction-corrected); (10) -- respective 1$\sigma$ uncertainty of the $(\text{V}-\text{I})_{M_I=-3.5}$; (11) -- the galaxy in our table is marked with * in this column, if the measured $(\text{V}-\text{I})_{M_I=-3.5}$ color is out of the range [1.22, 1.74] formally eligible for the color-metallicity relation (Equation \ref{formula:main-equation}); (12) -- the measured [Fe/H] value; (13) -- respective uncertainty of the [Fe/H]; (14) -- quality flag; a galaxy is marked with * in this column in the case of shallow photometric depth, meaning the photometric limit may affect the result, but the measurement is still available and considered reasonably reliable.

\startlongtable
\begin{deluxetable*}{cccccccccccccc}
\tablecaption{Results of our measurements of the metallicity [Fe/H]\label{tab:main}}
\tabletypesize{\scriptsize}
\tablehead{
\\[-3mm]
  \colhead{PGC} & \colhead{Name} & \colhead{Ra(2000)} & \colhead{Dec(2000)} & \colhead{D} & \colhead{$M_B$} & \colhead{T} & \colhead{E(B-V)} & \colhead{(V-I)} & \colhead{$1~\sigma$} & \colhead{$\notin[\dots]$} & \colhead{[Fe/H]} & \colhead{$1~\sigma$} & \colhead{Q}
}
\startdata
35&UGC12894&0.00625&39.4955&9.05&-13.45&10.0&0.108&1.032&0.043&*&-3.15&0.25&* \\
143&DDO221&0.03284&-15.46093&0.98&-14.09&9.0&0.038&1.368&0.024&&-1.58&0.08& \\
591&UGC64&0.12891&40.8759&8.16&-14.25&10.0&0.08&1.201&0.036&*&-2.27&0.17& \\
621&ESO349-031&0.13709&-34.57873&3.21&-11.87&10.0&0.012&1.286&0.021&&-1.89&0.08& \\
701&NGC24&0.16565&-24.96383&7.3&-17.66&5.0&0.02&1.422&0.023&&-1.4&0.08&* \\
1014&NGC55&0.24822&-39.19658&2.11&-18.76&8.0&0.013&1.499&0.024&&-1.17&0.06& \\
1034&NGC59&0.25698&-21.44455&4.89&-15.57&-3.0&0.022&1.397&0.013&&-1.48&0.04& \\
1038&ESO410-005&0.25868&-32.18061&1.93&-11.59&10.0&0.014&1.302&0.016&&-1.83&0.06& \\
1305&IC10&0.33975&59.2931&0.79&-16.39&10.0&1.556&1.554&0.013&&-1.03&0.03& \\
1641&ESO294-010&0.4426&-41.85558&2.03&-11.03&10.0&0.006&1.351&0.015&&-1.64&0.06& \\
1777&UGC288&0.48448&43.4308&6.84&-13.86&10.0&0.077&1.252&0.02&&-2.04&0.09& \\
2004&NGC147&0.55333&48.5086&0.75&-14.78&-3.0&0.172&1.52&0.042&&-1.11&0.11& \\
2329&NGC185&0.64937&48.3374&0.66&-14.91&-3.0&0.184&1.581&0.023&&-0.97&0.05& \\
2429&NGC205&0.6728&41.68538&0.8&-16.17&-2.0&0.085&1.51&0.027&&-1.14&0.07& \\
2555&NGC221&0.71161&40.86517&0.79&-16.05&-2.0&0.062&1.754&0.056&*&-0.69&0.06& \\
2557&M31&0.71232&41.26888&0.82&-21.4&3.0&0.062&2.099&0.111&*&-0.73&0.14& \\
2758&NGC247&0.78571&-20.76045&3.71&-18.56&7.0&0.018&1.552&0.022&&-1.03&0.05& \\
2789&NGC253&0.79253&-25.28844&3.7&-21.15&5.0&0.019&1.6&0.054&&-0.93&0.11& \\
2881&ESO540-030&0.82253&-18.0753&3.56&-11.49&10.0&0.023&1.298&0.013&&-1.85&0.05& \\
2902&DDO6&0.83047&-21.01302&3.43&-12.45&10.0&0.017&1.249&0.033&&-2.05&0.15& \\
2933&ESO540-032&0.84016&-19.90644&3.63&-11.45&10.0&0.02&1.305&0.01&&-1.82&0.04& \\
3238&NGC300&0.91487&-37.68448&2.09&-17.85&7.0&0.013&1.541&0.02&&-1.06&0.05& \\
3792&LGS3&1.06521&21.88973&0.65&-9.26&10.0&0.04&1.33&0.044&&-1.72&0.17&* \\
3844&IC1613&1.07994&2.11773&0.76&-14.62&10.0&0.025&1.373&0.054&&-1.56&0.19& \\
3974&UGC685&1.1229&16.68402&4.82&-14.44&9.0&0.057&1.392&0.022&&-1.49&0.08& \\
4126&NGC404&1.15749&35.71795&2.98&-16.43&-1.0&0.058&1.748&0.062&*&-0.7&0.07& \\
5818&M33&1.56414&30.65993&0.94&-18.98&6.0&0.042&1.57&0.037&&-0.99&0.08& \\
5896&NGC625&1.58475&-41.43676&4.01&-16.5&8.0&0.016&1.447&0.017&&-1.32&0.05& \\
6430&ESO245-05&1.75107&-43.59854&4.56&-15.68&9.0&0.016&1.212&0.018&*&-2.22&0.08&* \\
6699&UGC1281&1.82566&32.58622&5.27&-16.2&7.0&0.047&1.302&0.021&&-1.83&0.09& \\
6830&Phoenix&1.85174&-44.44474&0.44&-9.55&10.0&0.016&1.374&0.018&&-1.56&0.06& \\
7671&NGC784&2.02139&28.83724&5.38&-16.69&8.0&0.059&1.321&0.022&&-1.75&0.09& \\
8484&UGC1703&2.21531&32.8137&4.31&-11.59&-2.0&0.096&1.258&0.021&&-2.01&0.09& \\
9031&NGC891&2.37598&42.34712&9.97&-20.58&3.0&0.065&1.418&0.023&&-1.41&0.08& \\
9140&KKS03&2.41182&-73.51314&1.99&-10.72&-3.0&0.05&1.306&0.025&&-1.81&0.1& \\
9962&ESO115-21&2.62994&-61.34029&5.08&-15.58&7.0&0.026&1.349&0.022&&-1.65&0.08& \\
11139&ESO154-23&2.94727&-54.57064&5.87&-16.45&8.0&0.016&1.303&0.021&&-1.82&0.08& \\
11211&ESO199-007&2.96781&-49.38233&6.02&-12.55&10.0&0.022&1.189&0.031&*&-2.32&0.15& \\
11329&NGC1156&2.99514&25.2389&6.87&-17.86&8.0&0.224&1.279&0.016&&-1.92&0.07&* \\
12286&NGC1313&3.30433&-66.49759&4.32&-19.1&7.0&0.109&1.496&0.023&&-1.18&0.06& \\
12460&NGC1311&3.33528&-52.18582&5.54&-15.41&8.0&0.022&1.416&0.019&&-1.42&0.06& \\
12719&UGC2716&3.40207&17.75346&6.66&-15.24&8.0&0.138&1.24&0.019&&-2.09&0.08&* \\
13163&IC1959&3.55351&-50.41427&6.21&-16.07&8.0&0.011&1.398&0.021&&-1.47&0.07& \\
13294&PGC13294&3.59912&-45.19131&7.27&-13.23&10.0&0.012&1.28&0.023&&-1.92&0.1& \\
13826&IC342&3.78036&68.10139&3.56&-20.69&6.0&0.558&1.68&0.041&&-0.79&0.06&* \\
15345&NGC1569&4.51366&64.84816&3.19&-18.19&8.0&0.694&1.154&0.009&*&-2.49&0.04& \\
15439&UGCA92&4.53431&63.61611&3.22&-15.74&10.0&0.786&1.218&0.01&*&-2.19&0.05& \\
16389&ESO252-001&4.94963&-42.80068&6.97&-14.81&10.0&0.013&1.15&0.03&*&-2.51&0.15& \\
17302&ESO553-046&5.4516&-20.67785&6.7&-14.72&9.0&0.045&1.306&0.03&&-1.81&0.12& \\
18396&ESO364-029&6.09582&-33.08161&9.02&-16.31&9.0&0.045&1.053&0.02&*&-3.03&0.11&* \\
18536&NGC2188&6.16928&-34.1059&8.41&-17.78&8.0&0.033&1.189&0.012&*&-2.32&0.06&* \\
18731&ESO121-20&6.26509&-57.72557&6.07&-13.82&10.0&0.042&1.194&0.025&*&-2.3&0.12& \\
19041&ESO489-056&6.4382&-26.26578&6.32&-13.58&10.0&0.063&1.18&0.022&*&-2.37&0.11& \\
19337&ESO490-17&6.63252&-26.0002&6.34&-15.34&10.0&0.077&1.229&0.015&&-2.14&0.07& \\
19382&ESO308-022&6.65914&-40.72138&9.36&-14.14&10.0&0.09&1.211&0.015&*&-2.22&0.07&* \\
20125&FG202&7.08825&-58.51892&5.29&-14.18&10.0&0.12&1.187&0.017&*&-2.33&0.08& \\
21102&NGC2366&7.48169&69.21496&3.29&-16.11&9.0&0.036&1.296&0.029&&-1.85&0.12& \\
21199&ESO059-01&7.52179&-68.18753&4.63&-15.02&9.0&0.147&1.381&0.018&&-1.53&0.06& \\
21302&DDO44&7.56983&66.88399&3.2&-12.08&-3.0&0.041&1.367&0.016&&-1.58&0.06& \\
21396&NGC2403&7.61428&65.60242&3.19&-19.23&6.0&0.04&1.386&0.017&&-1.52&0.06& \\
21614&KK65&7.7088&16.56118&7.99&-14.31&10.0&0.032&1.368&0.02&&-1.58&0.07& \\
22277&UGC4115&7.9506&14.39101&7.88&-14.37&10.0&0.028&1.261&0.032&&-2.0&0.14& \\
23324&UGC4305&8.31788&70.72167&3.47&-16.76&9.0&0.032&1.298&0.018&&-1.85&0.07& \\
23344&ESO006-G001&8.32284&-85.14318&2.7&-12.58&10.0&0.192&1.324&0.016&&-1.74&0.06& \\
23521&M81DwA&8.39876&71.03082&3.42&-11.41&10.0&0.02&1.305&0.025&&-1.82&0.1& \\
24050&UGC4459&8.56868&66.17998&3.69&-13.44&10.0&0.038&1.32&0.014&&-1.76&0.05& \\
24213&UGC4483&8.61761&69.77639&3.58&-12.97&10.0&0.034&1.267&0.029&&-1.97&0.12& \\
26142&UGC4879&9.26742&52.84101&1.37&-11.95&9.0&0.016&1.374&0.014&&-1.56&0.05& \\
26259&NGC2835&9.29802&-22.35457&12.13&-20.07&5.0&0.087&1.455&0.057&&-1.29&0.17&* \\
26761&NGC2915&9.43662&-76.62644&4.28&-16.33&9.0&0.275&1.342&0.019&&-1.67&0.07& \\
27605&UGC5139&9.67564&71.1833&4.02&-14.59&10.0&0.05&1.31&0.023&&-1.8&0.09& \\
28120&NGC2976&9.78764&67.91626&3.67&-17.15&7.0&0.071&1.517&0.024&&-1.12&0.06& \\
28529&BK3N&9.89685&68.96914&4.17&-9.67&10.0&0.08&1.61&0.053&&-0.91&0.1&* \\
28630&M81&9.92587&69.06544&3.69&-20.96&3.0&0.08&1.862&0.185&*&-0.62&0.09& \\
28731&KDG61&9.95084&68.59203&3.67&-12.89&-1.0&0.073&1.391&0.012&&-1.5&0.04& \\
28757&UGC5336&9.95888&69.04596&3.86&-13.75&10.0&0.079&1.478&0.026&&-1.23&0.07& \\
28759&A0952+69&9.95856&69.27771&3.93&-11.54&10.0&0.084&0.844&0.065&*&-4.36&0.46&* \\
28826&d0958+66&9.98015&66.84977&3.83&-12.17&9.0&0.057&1.305&0.035&&-1.82&0.14& \\
28868&DDO69&9.99067&30.74639&0.74&-11.52&10.0&0.021&1.301&0.049&&-1.83&0.2& \\
28913&DDO70&10.00001&5.33222&1.43&-14.07&10.0&0.031&1.236&0.022&&-2.11&0.1& \\
29128&NGC3109&10.05191&-26.15949&1.34&-15.74&8.0&0.067&1.34&0.018&&-1.68&0.07& \\
29146&NGC3077&10.05517&68.73372&3.86&-17.62&9.0&0.067&1.736&0.041&&-0.71&0.05& \\
29194&Antlia&10.06777&-27.332&1.37&-9.83&10.0&0.078&1.296&0.024&&-1.85&0.1& \\
29231&BK5N&10.07808&68.25611&3.69&-10.56&-3.0&0.063&1.38&0.019&&-1.54&0.07& \\
29257&DDO71&10.08532&66.55905&3.64&-12.78&-3.0&0.098&1.364&0.027&&-1.59&0.1& \\
29388&KDG64&10.11717&67.82754&3.75&-12.6&-3.0&0.053&1.378&0.016&&-1.54&0.06& \\
29488&DDO74&10.14111&12.30639&0.26&-11.08&-1.0&0.036&1.314&0.025&&-1.78&0.1& \\
29653&DDO75&10.18355&-4.69282&1.44&-14.13&10.0&0.044&1.271&0.015&&-1.96&0.06& \\
29735&UGC5497&10.2133&64.10776&3.73&-12.34&9.0&0.02&1.314&0.017&&-1.78&0.07& \\
30819&IC2574&10.47349&68.41319&3.93&-16.78&8.0&0.036&1.413&0.014&&-1.42&0.04& \\
30997&DDO82&10.50967&70.62139&3.93&-14.67&9.0&0.041&1.398&0.021&&-1.47&0.07& \\
31286&BK6N&10.57515&66.00939&3.32&-10.75&-3.0&0.011&1.362&0.039&&-1.6&0.14& \\
31961&ESO376-016&10.72424&-37.04387&6.91&-13.92&10.0&0.058&1.026&0.037&*&-3.19&0.21&* \\
32249&NGC3377&10.7951&13.98564&10.44&-19.0&-2.0&0.03&1.765&0.046&*&-0.68&0.04& \\
32250&ESO318-013&10.79488&-38.85366&6.93&-14.56&8.0&0.077&1.214&0.02&*&-2.21&0.09& \\
32667&KDG73&10.882&69.54608&3.91&-10.88&10.0&0.019&1.298&0.051&&-1.85&0.21& \\
34176&DDO93&11.22447&22.1527&0.25&-9.43&-3.0&0.017&1.274&0.038&&-1.94&0.16&* \\
34554&NGC3621&11.30458&-32.81398&6.98&-19.77&7.0&0.081&1.409&0.029&&-1.44&0.09& \\
35286&UGC6456&11.46657&78.9939&4.63&-14.22&10.0&0.038&1.303&0.025&&-1.82&0.1& \\
35684&UGC6541&11.55801&49.23814&4.22&-13.77&10.0&0.019&1.273&0.022&&-1.95&0.09&* \\
35856&NGC3738&11.59665&54.52507&5.31&-16.82&9.0&0.01&1.369&0.019&&-1.58&0.07& \\
35878&NGC3741&11.60166&45.28561&3.22&-13.26&9.0&0.025&1.336&0.03&&-1.7&0.11& \\
36014&ESO320-14&11.63144&-39.22056&6.01&-13.66&10.0&0.143&1.252&0.03&&-2.04&0.13& \\
36758&UGC6757&11.7831&61.33505&4.61&-12.16&10.0&0.027&1.19&0.032&*&-2.32&0.15& \\
37050&DDO99&11.84801&38.88068&2.65&-13.53&10.0&0.026&1.23&0.025&&-2.13&0.11& \\
37369&ESO379-07&11.91199&-33.55876&5.45&-12.4&10.0&0.074&1.161&0.037&*&-2.46&0.18& \\
38148&NGC4068&12.06719&52.58843&4.38&-15.11&9.0&0.022&1.277&0.018&&-1.93&0.07& \\
38252&ESO379-024&12.08241&-35.74308&5.46&-12.59&10.0&0.074&1.144&0.032&*&-2.54&0.16& \\
38685&KUG1207+367&12.16565&36.43469&4.85&-13.04&9.0&0.025&1.361&0.037&&-1.6&0.13& \\
38688&NGC4144&12.16626&46.45729&6.9&-17.61&7.0&0.015&1.484&0.008&&-1.21&0.02& \\
38881&NGC4163&12.20256&36.16935&2.99&-13.84&9.0&0.02&1.339&0.016&&-1.69&0.06& \\
39023&NGC4190&12.22902&36.63325&2.83&-13.89&9.0&0.029&1.326&0.027&&-1.73&0.11& \\
39032&ESO321-014&12.23047&-38.23097&3.32&-12.8&10.0&0.094&1.189&0.037&*&-2.32&0.17& \\
39058&UGC7242&12.23567&66.09469&5.44&-14.16&10.0&0.019&1.382&0.019&&-1.53&0.07& \\
39145&DDO113&12.24947&36.21842&2.97&-11.6&-2.0&0.02&1.282&0.027&&-1.91&0.11& \\
39225&NGC4214&12.26088&36.32744&2.88&-17.16&8.0&0.022&1.43&0.011&&-1.37&0.03& \\
39228&CGCG269-049&12.26295&52.38729&4.62&-13.12&10.0&0.024&1.167&0.029&*&-2.43&0.14& \\
39316&UGC7298&12.27503&52.22773&4.19&-12.26&10.0&0.022&1.232&0.056&&-2.13&0.25&* \\
39422&NGC4244&12.29153&37.80716&4.3&-18.06&6.0&0.021&1.377&0.024&&-1.55&0.09& \\
39423&NGC4242&12.29171&45.61909&7.62&-17.95&7.0&0.012&1.446&0.017&&-1.32&0.05&* \\
39573&IC3104&12.3128&-79.72593&2.35&-14.91&9.0&0.41&1.289&0.015&&-1.88&0.06& \\
39600&NGC4258&12.31599&47.30382&7.65&-21.15&4.0&0.016&1.69&0.074&&-0.77&0.11& \\
39615&UGC7356&12.31924&47.09039&7.29&-13.64&-1.0&0.022&1.361&0.024&&-1.6&0.09& \\
39918&UGC7408&12.35423&45.81196&7.29&-15.81&9.0&0.012&1.197&0.024&*&-2.29&0.11& \\
40596&NGC4395&12.43025&33.54675&4.76&-17.88&8.0&0.017&1.496&0.026&&-1.18&0.07& \\
40665&UGCA281&12.43806&48.49417&5.69&-13.7&10.0&0.015&1.16&0.025&*&-2.46&0.12& \\
40750&MCG+09-20-182&12.44796&53.10536&6.12&-13.06&9.0&0.02&1.195&0.03&*&-2.3&0.14& \\
40791&DDO126&12.45138&37.14328&4.98&-14.42&10.0&0.014&1.217&0.03&*&-2.19&0.14&* \\
40904&DDO125&12.46158&43.49405&2.6&-14.22&9.0&0.02&1.314&0.024&&-1.78&0.09& \\
40973&NGC4449&12.46973&44.09303&4.26&-18.31&8.0&0.019&1.516&0.056&&-1.12&0.15& \\
41020&UGC7599&12.47447&37.23483&4.71&-13.46&10.0&0.017&1.207&0.031&*&-2.24&0.15& \\
41048&UGC7605&12.47742&35.71759&4.75&-13.68&10.0&0.014&1.091&0.04&*&-2.82&0.22&* \\
41066&NGC4455&12.47891&22.82148&6.61&-16.39&7.0&0.021&1.302&0.018&&-1.83&0.07& \\
41093&KK149&12.4812&42.17794&8.51&-14.75&10.0&0.024&1.279&0.026&&-1.92&0.11&* \\
41314&KK151&12.50658&42.90149&8.2&-13.85&9.0&0.02&1.236&0.021&&-2.11&0.1& \\
41636&UGC7698&12.54844&31.54103&4.87&-14.98&9.0&0.016&1.255&0.018&&-2.02&0.08& \\
42002&NGC4559&12.59936&27.95968&8.85&-20.01&6.0&0.018&1.322&0.022&&-1.75&0.09& \\
42134&PGC42134&12.62272&38.74379&6.15&-13.28&10.0&0.014&1.215&0.027&*&-2.2&0.12& \\
42275&UGCA292&12.64445&32.76688&3.85&-11.93&10.0&0.016&1.099&0.024&*&-2.78&0.13& \\
42408&NGC4605&12.6665&61.60903&5.54&-18.1&8.0&0.014&1.386&0.022&&-1.52&0.08& \\
42637&NGC4631&12.70219&32.54239&7.34&-20.65&7.0&0.017&1.404&0.017&&-1.45&0.06&* \\
42656&IC3687&12.7042&38.5026&4.58&-14.6&10.0&0.02&1.196&0.027&*&-2.29&0.12&* \\
42863&NGC4656&12.73267&32.17006&7.96&-19.05&8.0&0.013&1.145&0.019&*&-2.54&0.1&* \\
42936&ESO381-018&12.74515&-35.96658&5.44&-13.23&10.0&0.063&1.25&0.032&&-2.05&0.14& \\
43048&ESO381-20&12.76678&-33.83812&5.47&-14.53&10.0&0.066&1.199&0.027&*&-2.28&0.13& \\
43129&DDO147&12.78328&36.47642&3.0&-12.35&10.0&0.018&1.234&0.047&&-2.12&0.21& \\
43255&NGC4707&12.80637&51.1659&6.53&-15.7&10.0&0.011&1.249&0.014&&-2.05&0.06& \\
43495&NGC4736&12.8480500&41.1199900&4.41&-19.7357&2.0&0.0178&2.056&0.049&*&-0.68&0.05& \\
43869&NGC4789A&12.9015&27.14974&4.04&-14.0&10.0&0.009&1.181&0.02&*&-2.36&0.1& \\
43978&ESO443-09&12.91488&-28.34082&5.96&-12.15&10.0&0.064&1.163&0.026&*&-2.45&0.13& \\
44055&LEDA44055&12.9279&19.20923&4.83&-12.26&10.0&0.02&1.056&0.022&*&-3.01&0.13& \\
44182&M64&12.9455&21.68195&4.4&-19.53&2.0&0.041&2.312&0.028&*&-1.15&0.07& \\
44491&DDO155&12.97785&14.21735&2.19&-12.02&10.0&0.026&1.323&0.034&&-1.75&0.13& \\
44681&KK176&12.99905&-19.41139&7.28&-12.97&10.0&0.097&1.182&0.029&*&-2.36&0.14& \\
44982&UGCA319&13.03732&-17.23782&5.75&-14.12&10.0&0.081&1.249&0.022&&-2.05&0.1& \\
45084&DDO161&13.05466&-17.4229&6.03&-15.97&8.0&0.08&1.245&0.012&&-2.07&0.05& \\
45104&ESO269-37&13.0592200&-46.5868600&3.15&-11.77&10.0&0.1324&1.396&0.017&&-1.48&0.06& \\
45279&NGC4945&13.09092&-49.46791&3.47&-20.34&6.0&0.176&1.808&0.02&*&-0.65&0.01& \\
45314&IC4182&13.0971&37.60495&4.34&-16.23&8.0&0.014&1.316&0.019&&-1.77&0.08& \\
45372&UGC8201&13.10692&67.70693&4.83&-15.21&9.0&0.024&1.238&0.017&&-2.1&0.07& \\
45506&UGC8215&13.13432&46.82797&4.58&-12.31&10.0&0.011&1.325&0.021&&-1.74&0.08& \\
45546&UGC8245&13.14331&78.9371&4.73&-13.76&9.0&0.032&1.25&0.025&&-2.05&0.11& \\
45717&ESO269-058&13.17581&-46.99199&3.74&-15.04&9.0&0.109&1.378&0.015&&-1.54&0.05& \\
45916&ESO269-66&13.2191&-44.89&3.74&-13.68&-1.0&0.095&1.34&0.014&&-1.68&0.05& \\
45917&NGC5011C&13.2199&-43.26495&3.74&-14.15&-1.0&0.12&1.347&0.015&&-1.66&0.06& \\
45939&DDO167&13.22297&46.32028&4.24&-12.73&10.0&0.01&1.3&0.034&&-1.84&0.14&* \\
46039&DDO168&13.24116&45.91969&4.26&-15.24&10.0&0.015&1.288&0.026&&-1.89&0.11& \\
46127&UGC8331&13.25827&47.49963&4.41&-13.8&10.0&0.009&1.237&0.025&&-2.1&0.11& \\
46400&NGC5068&13.3153&-21.03919&5.16&-18.51&6.0&0.103&1.466&0.03&&-1.26&0.09& \\
46663&KK196&13.36307&-45.06245&3.96&-12.73&10.0&0.084&1.26&0.017&&-2.0&0.07& \\
46674&NGC5102&13.36601&-36.6301&3.67&-18.24&1.0&0.055&1.421&0.029&&-1.40&0.1& \\
46680&KK197&13.36719&-42.53474&3.84&-13.19&-3.0&0.155&1.471&0.025&&-1.25&0.07& \\
46957&NGC5128&13.42448&-43.01812&3.69&-20.74&-2.0&0.115&2.094&0.2&*&-0.73&0.24& \\
47073&IC4247&13.44568&-30.3625&5.18&-14.43&10.0&0.065&1.239&0.021&&-2.09&0.09& \\
47171&ESO324-24&13.46066&-41.47826&3.79&-15.49&8.0&0.112&1.302&0.025&&-1.83&0.1& \\
47404&M51&13.49797&47.19515&8.56&-21.26&5.0&0.035&1.772&0.04&*&-0.67&0.04& \\
47495&UGC8508&13.51233&54.91139&2.66&-13.07&10.0&0.015&1.268&0.029&&-1.97&0.12& \\
47762&NGC5206&13.56224&-48.15118&3.2&-16.43&-3.0&0.119&1.681&0.01&&-0.78&0.02& \\
47853&NGC5238&13.57847&51.61368&4.51&-14.77&10.0&0.01&1.301&0.017&&-1.83&0.07& \\
48029&ESO444-78&13.60856&-29.23536&5.42&-13.41&10.0&0.053&1.254&0.022&&-2.03&0.09& \\
48082&NGC5236&13.61691&-29.86556&4.89&-20.63&5.0&0.066&1.475&0.025&&-1.23&0.07& \\
48139&NGC5237&13.62752&-42.8469&3.32&-14.82&9.0&0.099&1.424&0.019&&-1.39&0.06& \\
48280&UGC8638&13.65543&24.77522&4.3&-13.75&10.0&0.013&1.315&0.018&&-1.78&0.07& \\
48332&UGC8651&13.66495&40.73914&3.11&-13.27&10.0&0.006&1.276&0.012&&-1.94&0.05& \\
48334&NGC5253&13.66555&-31.64006&3.44&-17.05&8.0&0.056&1.358&0.021&&-1.61&0.08& \\
48467&NGC5264&13.69354&-29.91272&4.78&-16.02&8.0&0.051&1.442&0.017&&-1.33&0.05& \\
48515&KK211&13.70157&-45.20517&3.68&-11.99&-2.0&0.111&1.444&0.022&&-1.33&0.07& \\
48738&ESO325-11&13.7502800&-41.8596500&3.40&-14.02&10&0.0880&1.257&0.016&&-2.02&0.07& \\
49050&ESO383-87&13.82152&-36.06337&3.19&-16.83&8.0&0.072&1.392&0.018&&-1.5&0.06& \\
49158&UGC8760&13.84754&38.02111&3.31&-13.22&10.0&0.016&1.293&0.016&&-1.86&0.06& \\
49448&UGC8837&13.91272&53.90331&7.23&-15.6&8.0&0.015&1.366&0.019&&-1.59&0.07& \\
49452&UGC8833&13.91347&35.83769&3.26&-12.31&10.0&0.012&1.264&0.017&&-1.98&0.07& \\
49615&ESO384-016&13.95039&-35.33292&4.49&-13.47&9.0&0.074&1.301&0.013&&-1.83&0.05& \\
50063&M101&14.0535&54.34906&6.65&-20.99&6.0&0.009&1.506&0.026&&-1.15&0.07& \\
50216&NGC5474&14.0837600&53.6622400&6.99&-17.79&8&0.0105&1.320&0.011&&-1.76&0.04& \\
50262&NGC5477&14.09258&54.46091&6.78&-15.01&9.0&0.011&1.26&0.024&&-2.0&0.1& \\
50961&UGC9128&14.26569&23.05556&2.3&-12.53&10.0&0.023&1.272&0.022&&-1.95&0.09& \\
51210&NGC5585&14.3300500&56.7294400&7.00&-18.2509&7&0.0157&1.297&0.013&&-1.85&0.05& \\
51472&DDO190&14.41209&44.52608&2.84&-14.21&9.0&0.012&1.213&0.012&*&-2.21&0.05& \\
51659&PGC51659&14.46766&-46.30517&3.62&-13.15&10.0&0.13&1.179&0.024&*&-2.37&0.11&* \\
52125&ESO222-010&14.58417&-49.42172&3.15&-13.29&10.0&0.27&1.241&0.022&&-2.09&0.1& \\
52142&UGC09405&14.59016&57.25567&5.8&-14.36&8.0&0.012&1.314&0.022&&-1.78&0.09& \\
52591&ESO272-25&14.72374&-44.70509&3.91&-14.26&10.0&0.164&1.34&0.008&&-1.68&0.03& \\
54392&ESO274-01&15.23706&-46.80911&2.79&-17.35&7.0&0.252&1.281&0.011&&-1.91&0.05& \\
60849&IC4662&17.78595&-64.6391&2.55&-15.61&9.0&0.07&1.481&0.016&&-1.22&0.05& \\
60921&NGC6503&17.82403&70.14433&6.26&-18.69&6.0&0.032&1.523&0.026&&-1.1&0.07& \\
61922&IC4710&18.47721&-66.98225&7.38&-17.21&8.0&0.089&1.368&0.022&&-1.58&0.08&* \\
62814&UGC11411&19.14511&70.2839&6.57&-14.05&9.0&0.119&1.339&0.029&&-1.69&0.11& \\
62836&NGC6744&19.16275&-63.8577&9.36&-21.26&4.0&0.043&1.623&0.025&&-0.88&0.05& \\
63000&NGC6789&19.27817&63.97242&3.55&-14.29&9.0&0.068&1.363&0.019&&-1.6&0.07&* \\
63287&ESO594-004&19.49973&-17.679&1.08&-11.57&10.0&0.124&1.162&0.021&*&-2.45&0.1& \\
63616&NGC6822&19.74916&-14.80034&0.51&-15.33&10.0&0.235&1.555&0.024&&-1.03&0.06& \\
65001&NGC6946&20.58127&60.15377&7.12&-21.41&6.0&0.342&1.505&0.019&&-1.15&0.05&* \\
65367&DDO210&20.78103&-12.84757&0.98&-11.17&10.0&0.05&1.256&0.039&&-2.02&0.17& \\
65603&IC5052&20.86808&-69.20142&5.5&-17.87&7.0&0.051&1.57&0.017&&-0.99&0.04& \\
67908&IC5152&22.04482&-51.29656&1.96&-15.56&9.0&0.025&1.421&0.03&&-1.4&0.1& \\
69468&KK258&22.67883&-30.79947&2.23&-10.51&-3.0&0.013&1.265&0.012&&-1.98&0.05& \\
69519&Tucana&22.6972&-64.42061&0.92&-9.26&-1.0&0.031&1.318&0.029&&-1.76&0.11& \\
71431&UGCA438&23.44107&-32.38858&2.21&-12.9&10.0&0.014&1.354&0.023&&-1.63&0.08& \\
71464&ESO347-G017&23.44896&-37.34695&8.42&-15.45&9.0&0.017&1.238&0.025&&-2.1&0.11&* \\
71538&Pegasus&23.47672&14.74278&0.97&-12.0&10.0&0.068&1.487&0.025&&-1.2&0.07& \\
71775&IC5332&23.5743&-36.10155&9.01&-18.9&7.0&0.017&1.322&0.031&&-1.75&0.12& \\
72675&ESO149-003&23.86748&-52.57736&7.0&-14.24&10.0&0.014&1.181&0.029&*&-2.36&0.14& \\
73049&NGC7793&23.96382&-32.59104&3.62&-18.34&6.0&0.019&1.371&0.025&&-1.57&0.09& \\
86635&AGC223254&12.46804&22.29083&6.32&-12.32&10.0&0.021&1.321&0.031&&-1.75&0.12& \\
86644&IC3840&12.86282&21.73575&5.99&-12.89&10.0&0.039&1.26&0.031&&-2.0&0.13& \\
87149&KK177&13.0449400&21.9979900&4.82&-11.93&-2&0.033&1.284&0.012&&-1.90&0.05& \\
95597&KKH37&6.79604&80.12395&3.43&-11.61&10.0&0.074&1.312&0.014&&-1.79&0.06& \\
166062&KK14&1.7452&27.28841&7.03&-12.08&10.0&0.072&1.448&0.027&&-1.31&0.08& \\
166064&KK16&1.92229&27.95376&5.63&-12.74&10.0&0.07&1.364&0.024&&-1.59&0.09& \\
166065&KK17&2.00281&28.83111&5.02&-11.54&10.0&0.055&1.346&0.029&&-1.66&0.11& \\
166073&KK27&3.35128&-66.32207&4.23&-11.96&10.0&0.083&1.318&0.018&&-1.77&0.07& \\
166077&KK35&3.75359&67.86466&4.1&-14.86&10.0&0.538&1.558&0.045&&-1.02&0.1&* \\
166084&CamB&4.88533&67.0995&3.49&-11.95&10.0&0.219&1.325&0.033&&-1.74&0.13&* \\
166101&KK77&9.83697&67.50918&3.79&-12.87&-3.0&0.138&1.379&0.03&&-1.54&0.11& \\
166127&NGC4258-DF6&12.31839&47.7305&7.54&-10.89&-1.0&0.016&1.328&0.042&&-1.73&0.16& \\
166130&KK135&12.3263&58.04395&5.46&-11.29&10.0&0.014&1.08&0.047&*&-2.88&0.26& \\
166135&AGC223231&12.38131&33.82909&7.99&-12.15&10.0&0.013&1.188&0.038&*&-2.33&0.18& \\
166137&KK144&12.42452&28.48244&5.99&-12.48&10.0&0.024&1.277&0.03&&-1.93&0.12& \\
166142&LEDA166142&12.73261&43.66156&4.33&-10.86&10.0&0.026&1.192&0.042&*&-2.31&0.2& \\
166146&KK166&12.82031&35.61272&4.38&-10.65&-3.0&0.015&1.321&0.04&&-1.75&0.16&* \\
166152&Cen6&13.08398&-40.08235&5.93&-12.98&10.0&0.102&1.168&0.033&*&-2.43&0.16& \\
166158&KK189&13.21256&-41.83206&4.23&-11.33&-3.0&0.111&1.335&0.021&&-1.7&0.08& \\
166159&KK191&13.22771&42.0436&8.3&-11.45&10.0&0.014&1.286&0.04&&-1.89&0.16&* \\
166163&KK195&13.3523&-31.52914&5.62&-11.82&10.0&0.061&1.092&0.044&*&-2.82&0.24& \\
166167&KK203&13.45778&-45.35242&3.78&-10.62&10.0&0.102&1.261&0.032&&-2.0&0.14& \\
166172&KK213&13.72675&-43.76933&3.76&-11.41&-3.0&0.098&1.366&0.039&&-1.59&0.14&* \\
166175&KK217&13.77143&-45.68481&3.49&-10.67&-3.0&0.121&1.482&0.029&&-1.21&0.08& \\
166176&KK218&13.7777&-29.97898&4.94&-10.4&-3.0&0.06&1.244&0.027&&-2.07&0.12& \\
166179&KK221&13.8129&-46.99664&3.81&-11.21&10.0&0.14&1.496&0.018&&-1.18&0.05& \\
166185&KK230&14.11957&35.06105&2.2&-9.28&10.0&0.014&1.193&0.051&*&-2.3&0.24& \\
385975&PGC385975&6.26902&-57.76422&5.94&-12.05&10.0&0.041&1.177&0.061&*&-2.38&0.3& \\
490287&ESO215-09&10.95833&-48.18386&5.47&-13.61&10.0&0.221&1.314&0.014&&-1.78&0.06& \\
592761&HIPASSJ1337-39&13.62374&-39.88871&5.08&-12.35&10.0&0.074&1.068&0.031&*&-2.95&0.17& \\
626659&LVJ2335-3713&23.58448&-37.2205&7.58&-12.83&10.0&0.016&1.534&0.04&&-1.08&0.1& \\
677373&DEEPJ1337-33&13.61674&-33.36252&4.55&-11.2&10.0&0.049&1.143&0.035&*&-2.55&0.18& \\
683190&HIP1133-32&11.55305&-32.96244&5.6&-12.73&10.0&0.062&1.254&0.034&&-2.03&0.15& \\
704814&2DFGRS-S431Z149&23.97798&-31.46746&3.66&-11.4&10.0&0.016&1.282&0.026&&-1.91&0.11& \\
1726175&AGC238890&13.54175&25.12361&5.3&-12.1&9.0&0.013&1.395&0.025&&-1.48&0.08& \\
2543081&LVJ1157+5638&11.96506&56.63797&8.75&-12.69&10.0&0.02&1.334&0.08&&-1.7&0.3& \\
2576103&MSB17J1343+58&13.71867&58.22758&5.59&-13.02&9.0&0.008&1.305&0.027&&-1.82&0.11& \\
2801026&KKR25&16.22992&54.37083&1.91&-9.44&-1.0&0.009&1.216&0.023&*&-2.2&0.1& \\
2807102&KKH5&1.12568&51.44045&5.3&-12.74&10.0&0.283&1.185&0.029&*&-2.34&0.14& \\
2807103&KKH6&1.58099&52.09172&3.38&-13.09&10.0&0.351&1.304&0.016&&-1.82&0.06& \\
2807114&KKH22&3.74908&72.0644&3.06&-12.11&10.0&0.385&1.213&0.019&*&-2.21&0.09& \\
2807133&KKH57&10.00356&63.18424&3.68&-10.05&-3.0&0.022&1.233&0.02&&-2.12&0.09& \\
2807150&KKH86&13.90931&4.24444&2.61&-10.32&10.0&0.027&1.237&0.031&&-2.1&0.14& \\
2807155&CasdSph&23.44215&50.67573&0.83&-11.76&-3.0&0.194&1.412&0.03&&-1.43&0.1& \\
2807157&KKH98&23.75945&38.71772&2.59&-10.89&10.0&0.123&1.38&0.036&&-1.54&0.12& \\
2815820&KKS53&13.18727&-38.9061&2.92&-11.26&-3.0&0.091&1.281&0.02&&-1.91&0.08& \\
2815822&KKs55&13.37018&-42.72865&3.85&-12.14&-3.0&0.143&1.324&0.015&&-1.74&0.06& \\
2815823&KKs57&13.69392&-42.58196&3.83&-10.81&-3.0&0.09&1.329&0.067&&-1.72&0.26& \\
2832120&BTS-151&12.72351&32.48225&7.6&-12.18&-2.0&0.016&1.339&0.036&&-1.69&0.14&* \\
3097113&HIPASSJ1351-47&13.85524&-46.9735&5.74&-11.91&10.0&0.143&1.159&0.027&*&-2.47&0.14& \\
3097691&Cetus&0.43634&-11.04785&0.79&-10.2&-1.0&0.029&1.352&0.029&&-1.64&0.11& \\
3097727&ScdE1&0.3977&-24.70498&4.28&-10.5&-3.0&0.015&1.272&0.013&&-1.95&0.06& \\
3097827&F8D1&9.74642&67.43861&3.75&-11.46&-3.0&0.108&1.553&0.02&&-1.03&0.05& \\
3097828&FM1&9.75278&68.76497&3.78&-11.86&-3.0&0.078&1.344&0.016&&-1.67&0.06& \\
3994669&HIPASSJ1247-77&12.79272&-77.58262&3.47&-12.9&10.0&0.757&1.02&0.034&*&-3.22&0.2& \\
4074723&SDSS-J124029.91+472203.9&12.67498&47.36776&7.63&-11.5&10.0&0.017&1.179&0.05&*&-2.37&0.24&* \\
4320422&LVJ1218+4655&12.30307&46.9169&8.26&-12.84&8.0&0.019&1.156&0.042&*&-2.48&0.21&* \\
4343356&AGC742601&12.8269&21.91818&6.47&-11.88&10.0&0.034&1.416&0.038&&-1.42&0.12& \\
4608690&ANDROMEDA-17&0.61861&44.32222&0.88&-7.08&-2.0&0.074&1.238&0.031&&-2.1&0.14&* \\
4614882&HIPASSJ1348-37&13.80931&-37.96812&5.66&-11.49&10.0&0.078&1.208&0.031&*&-2.23&0.14& \\
4689184&KK242&17.88011&70.13733&6.46&&&0.036&1.264&0.049&&-1.99&0.21&* \\
4689187&CenN&13.80256&-47.56495&3.66&-10.93&-3.0&0.14&1.336&0.016&&-1.7&0.06& \\
4689195&IKN&10.13499&68.39926&3.75&-13.67&-3.0&0.058&1.411&0.027&&-1.43&0.09& \\
4689216&HS117&10.357&71.11615&3.95&-11.6&10.0&0.115&1.294&0.021&&-1.86&0.09& \\
4689222&ANDROMEDA-9&0.88139&43.19583&0.74&-8.21&-3.0&0.075&1.321&0.033&&-1.75&0.13& \\
5056918&ANDROMEDA-18&0.03736&45.08889&1.31&-8.73&-3.0&0.106&1.27&0.014&&-1.96&0.06& \\
5056920&ANDROMEDA-20&0.12519&35.13233&0.86&-5.77&-3.0&0.059&1.297&0.057&&-1.85&0.23&* \\
5056921&ANDROMEDA-10&1.10936&44.80439&0.74&-6.56&-3.0&0.129&1.182&0.031&*&-2.36&0.15& \\
5056922&ANDROMEDA-14&0.85972&29.69694&0.88&-7.78&-3.0&0.06&1.282&0.059&&-1.91&0.25&* \\
5056932&d0939+71&9.6544&71.31165&3.65&-8.38&10.0&0.038&1.228&0.038&&-2.14&0.17& \\
5056933&d0944+69&9.73959&69.21105&3.84&-7.33&-3.0&0.102&1.242&0.049&&-2.08&0.21&* \\
5056934&d0955+70&9.92043&70.40802&3.45&-8.83&-3.0&0.15&1.417&0.054&&-1.41&0.17&* \\
5056936&d0959+68&9.99254&68.65691&4.27&-10.47&10.0&0.073&1.416&0.041&&-1.42&0.13& \\
5056937&d1006+67&10.11284&67.20113&3.62&-8.4&-3.0&0.071&1.478&0.089&&-1.23&0.25&* \\
5056941&d1028+70&10.4777&70.23361&3.84&-11.89&10.0&0.036&1.332&0.025&&-1.71&0.1& \\
5056942&d1041+70&10.688&70.15096&3.69&-8.53&-2.0&0.061&1.411&0.054&&-1.43&0.18& \\
5056943&d0926+70&9.44108&70.50665&3.4&-9.65&10.0&0.207&1.192&0.04&*&-2.31&0.19&* \\
5056944&d0944+71&9.74289&71.48266&3.46&-11.14&-1.0&0.039&1.332&0.018&&-1.71&0.07& \\
5056947&d1015+69&10.25192&69.03753&4.07&-7.76&-1.0&0.049&1.221&0.085&&-2.18&0.39&* \\
5056993&LVJ1243+4127&12.73213&41.45683&4.82&-11.29&10.0&0.018&1.336&0.017&&-1.7&0.07& \\
5057032&DDO169NW&13.25558&47.54361&4.32&-10.22&10.0&0.008&1.265&0.017&&-1.98&0.07& \\
5057055&AGC112521&1.68552&27.32255&6.07&-11.39&10.0&0.061&1.243&0.045&&-2.08&0.2&* \\
5057226&ANDROMEDA-23&1.48939&38.71889&0.83&-9.0&-3.0&0.066&1.261&0.038&&-2.0&0.16& \\
5057227&ANDROMEDA-24&1.30833&46.36611&0.82&-6.94&-3.0&0.083&1.192&0.035&*&-2.31&0.17&* \\
5057228&ANDROMEDA-25&0.50247&46.85194&0.8&-8.57&-3.0&0.101&1.28&0.034&&-1.92&0.14&* \\
5057229&ANDROMEDA-26&0.396&47.91611&0.83&-5.17&-3.0&0.11&1.253&0.055&&-2.03&0.24&* \\
5057231&ANDROMEDA-21&23.91325&42.47083&0.82&-8.47&-3.0&0.093&1.303&0.036&&-1.82&0.14&* \\
5057232&ANDROMEDA-22&1.46111&28.09028&0.99&-5.83&-3.0&0.079&1.292&0.048&&-1.87&0.2&* \\
5059815&AGC748778&0.10953&15.51083&6.23&-10.35&10.0&0.064&1.318&0.051&&-1.77&0.2& \\
5060050&AGC174585&7.60286&9.98639&7.72&-11.71&10.0&0.038&1.177&0.042&*&-2.38&0.21&* \\
5060430&ANDROMEDA-29&23.98211&30.75556&0.82&-7.52&-2.0&0.046&1.338&0.036&&-1.69&0.14&* \\
5060431&ALFAZOAJ1952+1428&19.86994&14.47333&8.39&-13.82&9.0&0.26&1.288&0.02&&-1.89&0.08& \\
5065056&ANDROMEDA-30&0.60965&49.64681&0.68&-7.48&-3.0&0.166&1.208&0.037&*&-2.23&0.17& \\
5065677&ANDROMEDA-31&22.97119&41.29111&0.84&-10.97&-3.0&0.14&1.256&0.025&&-2.02&0.11& \\
5065678&ANDROMEDA-32&0.59983&51.55972&0.83&-11.53&-3.0&0.195&1.382&0.025&&-1.53&0.09& \\
5067061&ANDROMEDA-33&3.02322&40.98833&0.82&-9.62&-2.0&0.13&1.259&0.029&&-2.01&0.12& \\
5067392&M101-DwA&14.11383&53.74133&6.58&-9.33&-2.0&0.01&1.339&0.044&&-1.69&0.17& \\
5072213&CENA-DW1&13.50393&-41.89178&3.52&-12.1&-2.0&0.127&1.382&0.021&&-1.53&0.07& \\
5072214&CENA-DW2&13.49926&-41.87294&3.79&-8.04&-2.0&0.13&1.24&0.032&&-2.09&0.14& \\
5072710&PiscesA&0.24612&10.81307&5.36&-10.7&10.0&0.098&1.104&0.04&*&-2.75&0.21& \\
5072711&PiscesB&1.31992&11.12164&8.82&-12.16&10.0&0.056&1.1&0.029&*&-2.77&0.16&* \\
5098252&AntliaB&9.81558&-25.99&1.43&-9.6&10.0&0.08&1.261&0.024&&-2.0&0.1& \\
5509262&NGC5128-S1&13.50597&-42.1925&3.61&-11.94&-2.0&0.112&1.804&0.086&*&-0.65&0.06&* \\
5509263&CENA-DW4&13.38397&-41.78611&3.97&-10.0&-2.0&0.126&1.307&0.04&&-1.81&0.16& \\
5509264&CENA-DW5&13.33122&-41.99361&3.84&-6.6&-2.0&0.123&1.166&0.061&*&-2.43&0.3& \\
5509265&CENA-DW6&13.43267&-41.09417&3.57&-8.51&-2.0&0.112&1.376&0.057&&-1.55&0.2& \\
5509266&CENA-DW7&13.44131&-43.55667&3.65&-8.35&-2.0&0.115&1.987&0.384&*&-0.63&0.35&* \\
5808780&AGC229379&12.50944&23.20556&6.88&-10.74&10.0&0.018&1.347&0.062&&-1.66&0.23& \\
6726347&LBTJ115205.6+544732.2&11.86822&54.79229&5.96&-10.64&10.0&0.009&1.248&0.038&&-2.05&0.17& \\
6740710&WOC2017-7&0.91695&-23.16914&3.62&-9.76&10.0&0.016&1.214&0.024&*&-2.21&0.11&
\enddata
\end{deluxetable*}

\begin{figure}[h]
    \center
    \includegraphics[width=0.49\textwidth]{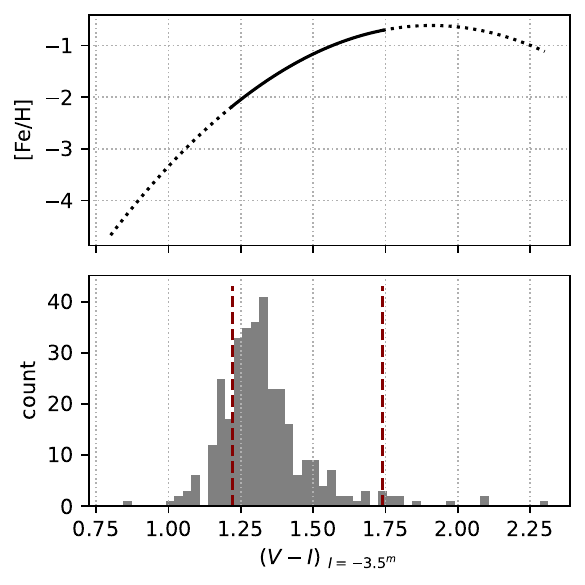}
    \includegraphics[width=0.49\textwidth]{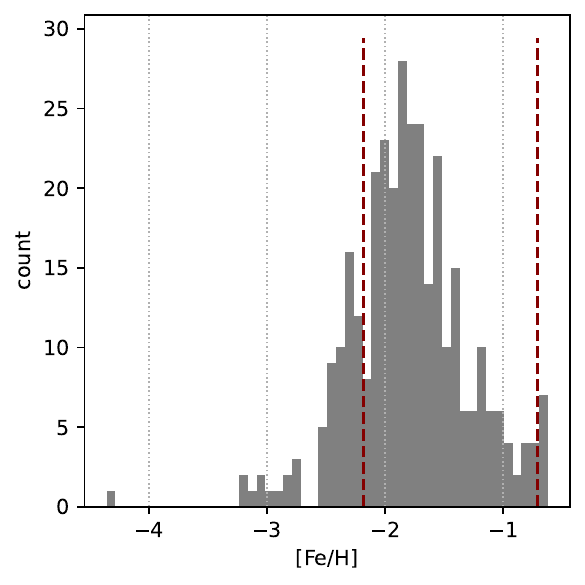}
    \caption{\textbf{Top left}: Metallicity versus color index plot based on equation (\ref{formula:main-equation}). The solid line indicates the color range where the equation was originally determined, while the dotted line represents an extrapolation beyond this range. \textbf{Bottom left}: Histogram of galaxy distribution as a function of the derived color index. Vertical lines denote the respective limits of the equation; approximately 77\% of the measurements fall within these bounds. \textbf{Right}: Histogram of galaxy distribution as a function of metallicity. Vertical lines indicate the range of reliable measurements; values outside this range result from the extrapolation.}
    \label{fig:feh_histplot}
\end{figure}

The histogram of color and metallicity distribution of galaxies in our sample is depicted in Fig.~\ref{fig:feh_histplot}. The upper left panel shows the $(\text{V}-\text{I})_{-3.5}$--metallicity relation as found by \citet{1993ApJ...417..553L}. The solid line highlights the interval corresponding to the limits of the color indices of the globular clusters used for this calibration. It is clearly visible that the overwhelming majority of galaxies in our sample fall within the required color index range. The peak of the distribution is approximately at $(\text{V}-\text{I})_{-3.5}=1.31^m$. There are a few galaxies with a color index redder than 1.74, and a number of objects with an average RGB color bluer than 1.22 appear, with the color of individual objects close to ~1. The respective distribution of the measured metallicities of the RGB stars appears at the right panel of the Fig.~\ref{fig:feh_histplot}. Here we can see the low metallicities typical of dwarf galaxies, with the peak of the distribution being at about $-1.89\pm0.03$~dex. The ``red tail" of the color distribution falls to the last bin of the histogram, and the group of galaxies with the most blue color gives us a number of objects with extremely metal-poor old stellar population. Although the low-metallicity tail looks quite reasonable for galaxies with an RGB color of about 1.1 to 1.2, the metallicity of a separate group of the bluest RGBs remains questionable.

\begin{figure}[h]
    \center
    \includegraphics[width=0.49\textwidth]{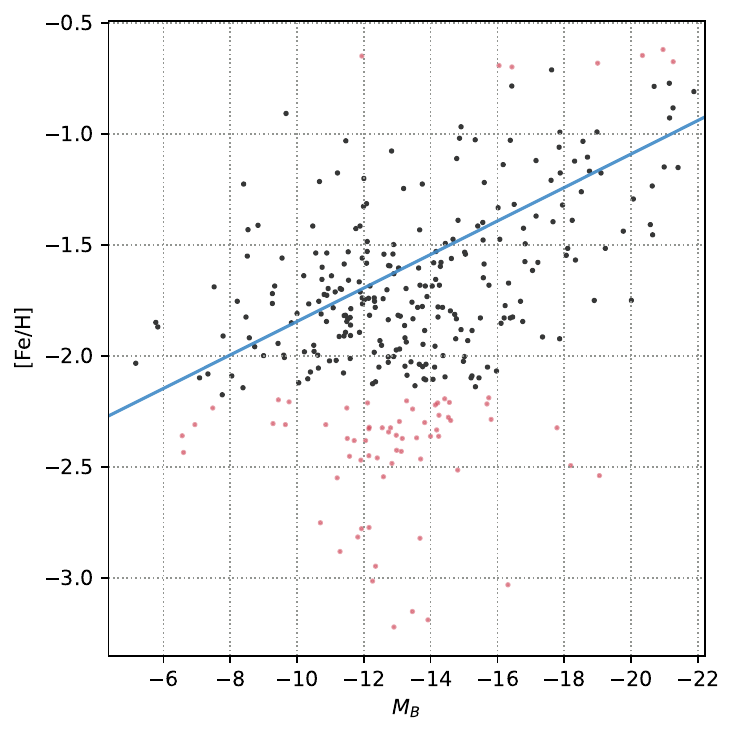}
    \includegraphics[width=0.49\textwidth]{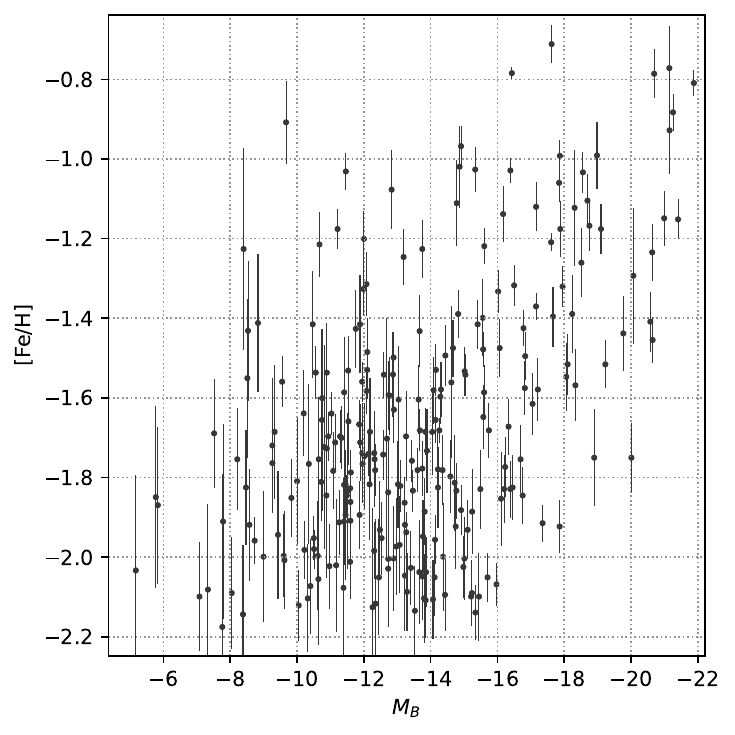}
    \caption{\textbf{Left}: Relation between galaxy metallicity and absolute magnitude $M_B$. Black points show galaxies within the applicability range of relation (\ref{formula:main-equation}), while red points indicate those outside it. The solid line represents a linear regression fit that accounts for measurement uncertainties. The best-fit relation is given by $\text{[Fe/H]} = -2.6 - 0.075 \, M_B$. Expressed in terms of luminosity, this corresponds to $Z \propto L_B^{0.19}$. \textbf{Right}: The same relation, but including only galaxies within the [1.22, 1.74] formally eligible color range, shown with confidence intervals. Error bars indicate the uncertainties in the metallicity estimates.}
    \label{fig:mb-feh_distribuition}
    \label{fig:mb-feh_with_errorbars}
\end{figure}

In the following Fig.~\ref{fig:mb-feh_distribuition} the measured [Fe/H] versus the total absolute B magnitude is presented, according to our Table~\ref{tab:main}. 
It can be seen that the range of absolute magnitudes of the studied galaxies is quite wide. Our sample mainly consists of dwarf galaxies with $M_B$ fainter than $-18^m$, including a few very faint objects with $M_B$ fainter than $-7^m$, 
and also presents measurements of the RGB metallicity of a subsample of nearby giant galaxies with $M_B$ brighter than $-18^m$. 
This subsample is small, including 36 galaxies, and we can nevertheless see that the metallicities we determined for the old stellar populations of giant galaxies reliably follow the absolute magnitude--metallicity relationship for the nearby dwarf galaxies in our sample.
It should be noted here that the ACS (or WFPC2) field of view is usually smaller than the size of the objects in this subsample, and the camera position was mainly chosen to avoid crowded stellar fields typical of the central regions of galaxies. Therefore, in a few cases our [Fe/H] values belong
to the outskirts of the bright galaxies. However, these
objects do not stand out from the others.
Although the $M_B$--$\text{[Fe/H]}$ dependence has a substantial scatter (Fig.~\ref{fig:mb-feh_distribuition}), the correlation of the given values is clearly visible. The solid line represents a linear regression fit that accounts for measurement uncertainties. The best-fit relation is given by $\text{[Fe/H]} = -2.6 - 0.075 \, M_B$. Expressed in terms of luminosity, this corresponds to $Z \propto L_B^{0.19}$. The right panel of Fig.~\ref{fig:mb-feh_with_errorbars} demonstrate the same relation with the uncertainties of the estimated [Fe/H] for clarity. 

\begin{figure}[h]
    \center
    \includegraphics[width=0.49\textwidth]{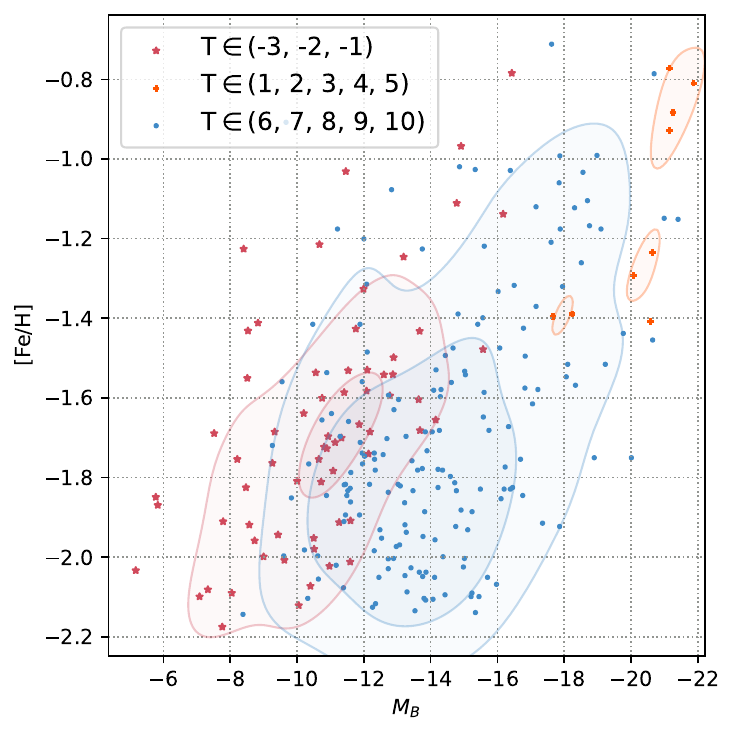}
    \includegraphics[width=0.49\textwidth]{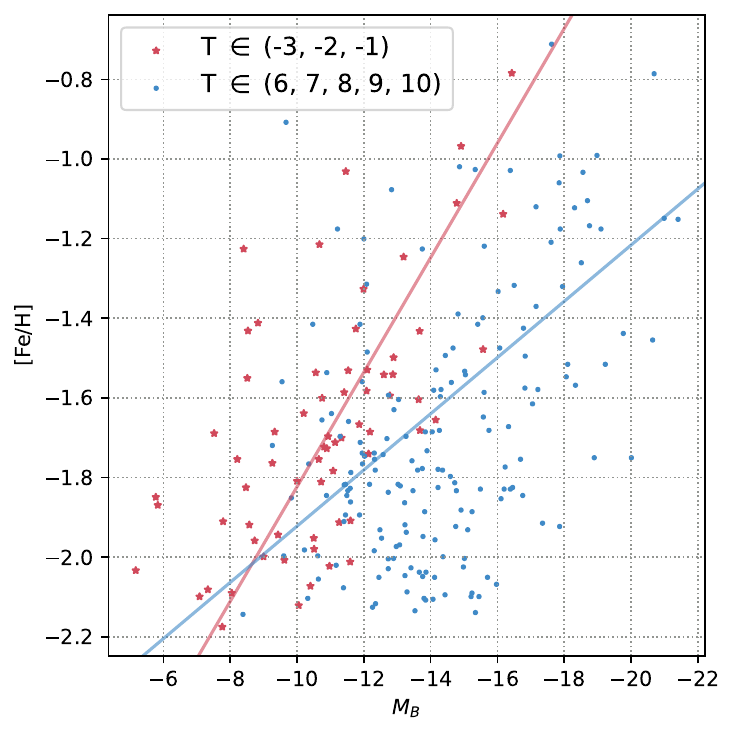}
    \caption{Distribution of sample galaxies by morphological type on the $M_B$–metallicity plane. Only galaxies with color indices within [1.22, 1.74] range are included. Morphological types according to de Vaucouleurs are grouped and color-coded as follows: “very early” types ($T < 0$) in red, “very late” types ($T > 5$) in blue, and intermediate types ($0 \leq T \leq 5$) in orange. \textbf{Left}: Individual galaxies with overlaid isodensity contours (in the same colors), highlighting regions of higher concentration. \textbf{Right}: Linear regression fits shown separately for “very early” and “very late” types. The best-fit relations are $\text{[Fe/H]} = -3.26 - 0.144 \, M_B$ (red line) and $\text{[Fe/H]} = -2.63 - 0.071 \, M_B$ (blue line), respectively. Intermediate types are few in number and therefore not included in the fit.}

    \label{fig:mb-feh_by_types}
\end{figure}

\begin{figure}[h]
    \center
    \includegraphics[width=0.49\textwidth]{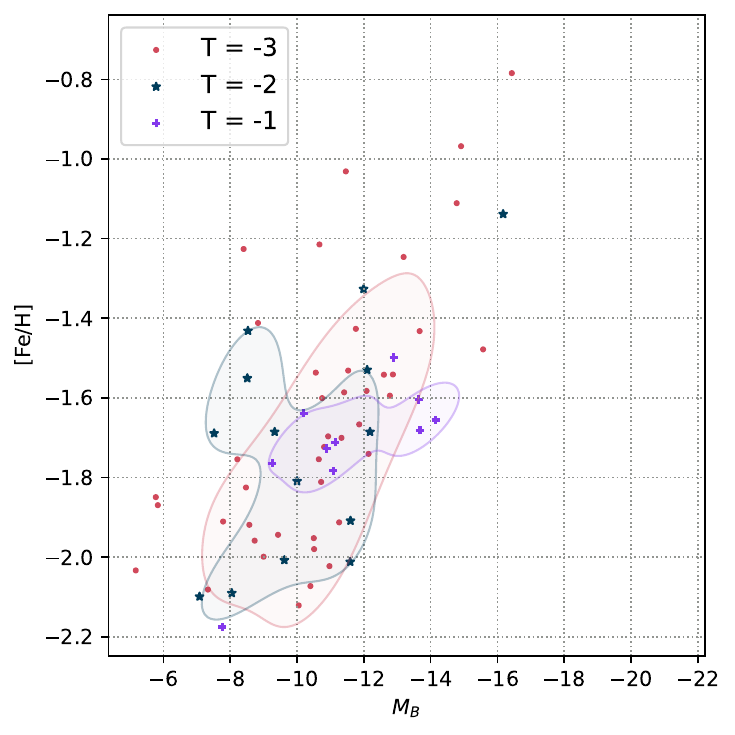}
    \includegraphics[width=0.49\textwidth]{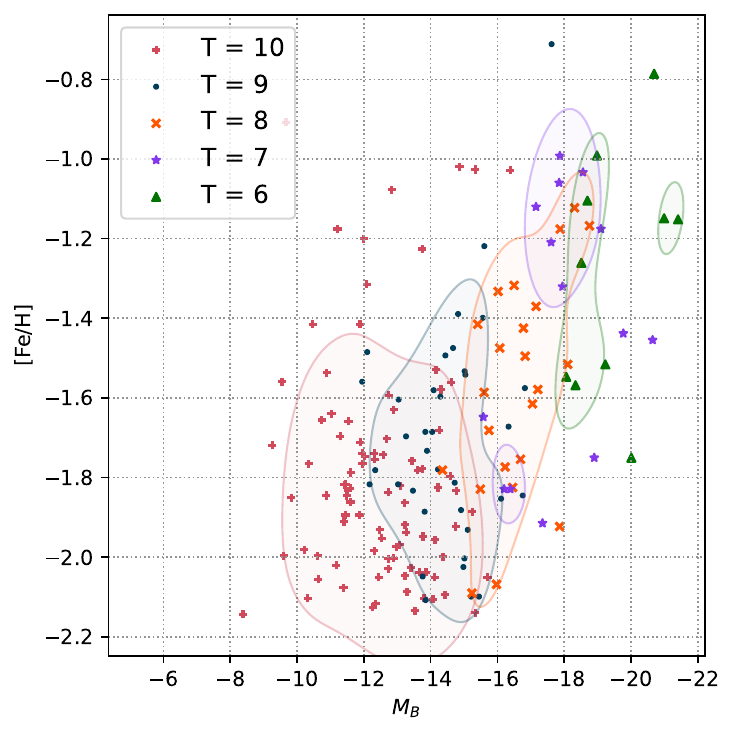}
    \caption{Total absolute magnitude $M_B$ versus metallicity. Only galaxies with color indices within the validity range were included. Galaxies are color-coded according to their de Vaucouleurs morphological types. Early-type galaxies are shown in the left panel, and all other types in the right panel. Isodensity contours are added, highlighting regions of higher concentration. It is noteworthy that the late-type galaxies form a rather well-defined sequence in the diagram.}
    \label{fig:mb-feh_by_exact_types}
\end{figure}

Here we can see growth of the average metallicity of the red giant stars with an increase of the total absolute magnitude of the galaxy.
This relation appears to be quite natural, given the known fact of greater intensity and longer periods of star formation is typical for brighter galaxies, and, therefore, we can expect higher overall metallicity of these galaxies.
The similar ratio was considered in the work of \citet{2008MNRAS.384.1544S} (see the comparison below). One of 
the obvious reasons for this luminosity-metallicity relation (and mass-metallicity relation) could be star formation efficiency: Massive galaxies evolve quickly, converting gas efficiently into stars, increasing metallicity, whereas low-mass galaxies evolve slowly, have higher gas fractions, and thus diluted metallicity.

It should be noted, that most mass-- and luminosity--metallicity relations were established from the metallicity measurements of gas-phase spectral abundance, giving tighter relations (see, for example, the historical work of \citet{1986ApJ...303...39D}).
At the same time, a number of studies indicate that the LZR for nearby galaxies (including dwarf ones) could be a combination of different mechanisms, predominantly of star formation efficiency and gravitational retention of metals, in the sense that massive galaxies have deeper gravitational wells that retain metals against loss from feedback processes like supernova and stellar winds, while low-mass galaxies lose more metals (see, for example \citet{2007ApJ...658..941D} and \citet{2020ApJ...891..181M}).
In recent years, the superior observational possibilities allow to measure a metallicity of individual stars in nearby galaxies, provided us with the accurate data for further study of the LMR (see, for example, \citet{2013ApJ...779..102K}).

The distribution of our sample galaxies by morphological type on the $M_B$-metallicity plane is shown in Fig.~\ref{fig:mb-feh_by_types}. It is obvious that the vast majority of our sample are dwarf galaxies, which can be generally divided into dwarf spheroidals (early types) and dwarf irregulars (late types) (see the left panel of Fig.~\ref{fig:mb-feh_by_types}). Note that the two subsamples are separated by their position in the plot. The right panel demonstrate the respective regressions for these two subsamples. It is clearly visible that the metallicity of early type galaxies is higher on average, and the slopes are similar for the early and the late type galaxies. It is worth noting here that the dwarf and giant galaxies in our sample form a single sequence in the absolute magnitude-metallicity plane, although the subsample of bright galaxies is small. 

The distribution of galaxies in our sample by distinct morphological types using various symbols is demonstrated in Fig.~\ref{fig:mb-feh_by_exact_types}. Whereas the early types are distributed quite uniformly (the left panel), the distribution areas corresponding to the different late types (the right panel of the Figure~\ref{fig:mb-feh_by_exact_types}) form a rather curious sequence: the "earlier" the morphological type of the object, the brighter and more metal rich it is.

\section{Comparison with previous studies}

\begin{figure}[h]
    \center
    \includegraphics[width=0.4\textwidth]{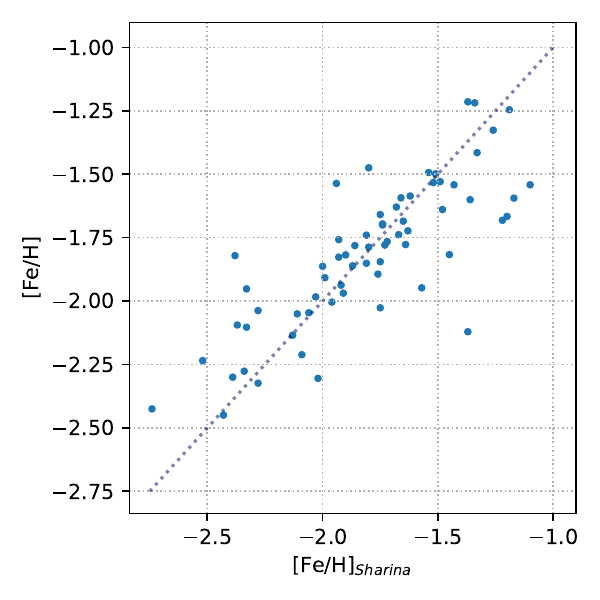}
    \includegraphics[width=0.4\textwidth]{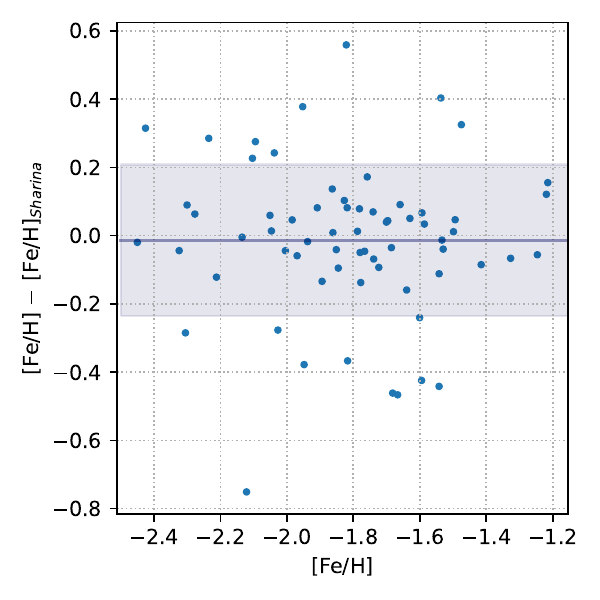}
    \caption{\textbf{Left}: Comparison of our metallicity measurements with those reported by \citet{2008MNRAS.384.1544S}. The x-axis shows Sharina’s values, while the y-axis shows our results, for a total of 67 galaxies. The dashed blue line indicates the one-to-one relation ($x = y$). \textbf{Right}: Distribution of the differences between the two datasets. The x-axis shows our measurements, and the y-axis shows the difference. The horizontal blue line marks the mean offset ($-0.013$ dex), and the shaded band represents the range of one standard deviation around the mean ($\pm 0.222$).}
    \label{fig:comparison_sharina}
\end{figure}

\begin{figure}[h]
    \center
    \includegraphics[width=0.4\textwidth]{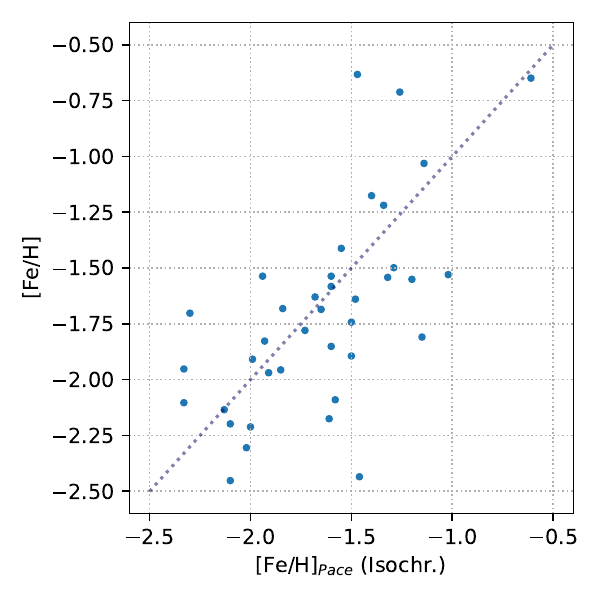}
    \includegraphics[width=0.4\textwidth]{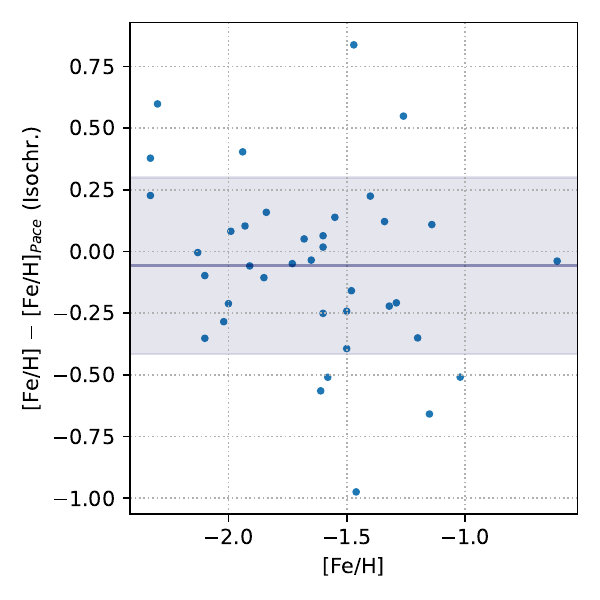}
    
    \includegraphics[width=0.4\textwidth]{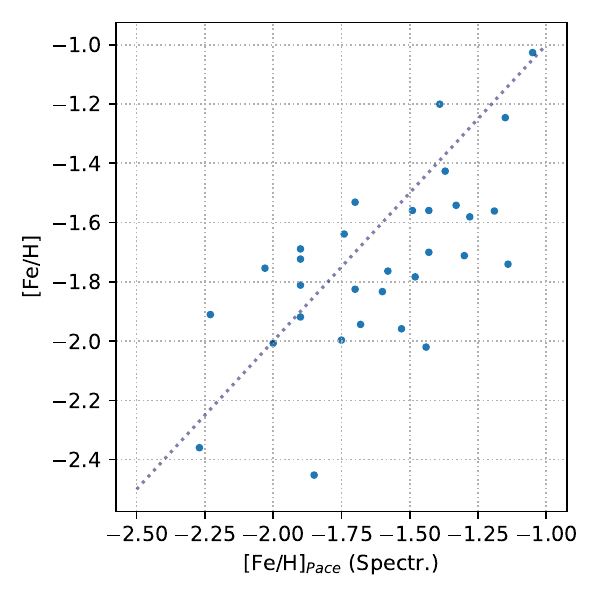}
    \includegraphics[width=0.4\textwidth]{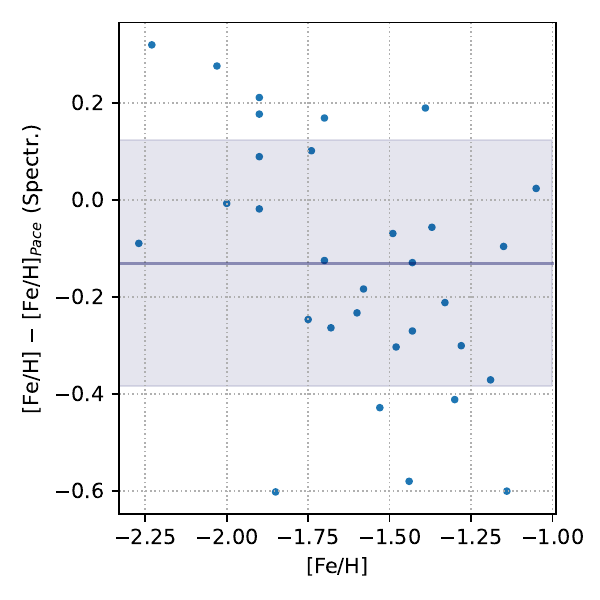}
    \caption{Comparison of our metallicity measurements with those reported by \citet{2025OJAp....8E.142P}. \textbf{Top}: Results obtained using the isochrone–fitting method ($38$ galaxies). \textbf{Bottom}: Results based on spectroscopic measurements ($31$ galaxies). In both panels, the plots are presented in the same format as in Figure \ref{fig:comparison_sharina}: on the left, a direct comparison with the dashed blue line indicating $x=y$; on the right, the distribution of differences. The mean offsets are $-0.059$ and $-0.130$ dex, with standard deviations $0.358$ and $0.254$, respectively (shown by the horizontal blue line and shaded $\pm1\sigma$ region).}
    \label{fig:comprison_pace}
\end{figure}

To assess the reliability of our metallicity measurements, we compared our results with those obtained in previous studies. 
A fairly large sample of Local Volume galaxies was analyzed in the work of \citet{2008MNRAS.384.1544S}. An approach similar to ours was used in this study, i.e. a metallicity was determined from the average RGB color of nearby galaxies obtained with HST/WFPC2/ACS.
For 67 galaxies in common between the two samples, we constructed a direct comparison of the metallicities as well as the distribution of the differences (Fig.~\ref{fig:comparison_sharina}). The results show good overall agreement, with an average offset of only $-0.013$ dex and a scatter consistent with the expected measurement uncertainties.

We also compared our results with the values from the recent paper of \citet{2025OJAp....8E.142P}, where a compilative catalog of the observed properties of dwarf galaxies within about 3 Mpc is presented. The comparison was made with the [Fe/H] values obtained via isochrone fitting ($38$ galaxies) and spectroscopic measurements ($31$ galaxies) (Fig.~\ref{fig:comprison_pace}). The mean offsets are $-0.059$ and $-0.130$ dex, with dispersions $0.358$ and $0.254$, respectively. Agreement is generally good, though the isochrone method shows larger scatter, while spectroscopy displaying a larger offset.

\section{Conclusions}
The Extragalactic Distance Database (EDD: http://edd.ifa.hawaii.edu/) hosts the CMDs/TRGB catalog, 
which provides uniformly measured, high-precision TRGB distances on a common scale, along with tables of PSF photometry for resolved stars in more than 500 Local Volume galaxies \citep{2021AJ....162...80A} observed with HST. We exploit the remarkable capabilities of this catalog to estimate the metallicity of the old stellar populations uniformly using
resolved red giant branch of the LV galaxies. We use the  technique of \citet{1993ApJ...417..553L} to measure homogeneously the metallicity [Fe/H] for the first time for this LV sample, which is almost complete up to absolute magnitude --12 mag. We summarize the results of our study as follows.

\begin{itemize}
    \item The color index and metallicity were successfully measured for 445 galaxies out of 558, the remaining objects were excluded because they were too far to estimate TRGB distance, or the photometric limit cuts off the red part of the RGB, preventing the color index from being estimated. Then, 111 measurements from the remaining ones were rejected, because at the level of $M_I = –3.5^m$ the photometric limit for these objects could significantly bias the results. Therefore, our resulting table contains 334 galaxies.
    \item To estimate the (V-I) color of the red giant branch (RGB) at $M_I = -3.5$ for measuring galaxy metallicity [Fe/H], we used the maximum stellar density in the ($M_I, V-I$) diagram, smoothed with a Gaussian kernel and refined via Monte Carlo simulations (500-1000 realizations) accounting for photometric and distance errors. This yields a reliable color with ~0.03 mag uncertainty on average, though sparser RGBs required parabolic fitting, and Milky Way contamination was handled by simulating foreground stars with TRILEGAL and manual isolation. The procedure is implemented in a public Python3 package with a Qt5 GUI for interactive photometric data processing, outputting PDF reports and JSON results.
    \item Most of the galaxies in our sample fall in the color range of [1.22, 1.74] of the original calibration \citep{1993ApJ...417..553L}, and respective low metallicities are typical of dwarf galaxies. The resulting distribution of the measured metallicity has a peak at [Fe/H] = $-1.89\pm0.03$ dex. A small group of galaxies with the most blue colors gives us a number of objects with extremely metal-poor old stellar populations.
    \item Our sample covers quite a wide absolute magnitude range of of galaxies, including a few very faint objects with $M_B$ fainter than $-7^m$ and a subsample of nearby giant galaxies with $M_B$ brighter than $-18^m$. We found the clear positive correlation of [Fe/H] with $M_B$. The best-fit linear regression is given by the following relation: $\text{[Fe/H]} = -2.6 - 0.075 \, M_B$.
    \item The subsample of the nearby giant galaxies follows the same $M_B$--$\text{[Fe/H]}$ relation as the dwarf ones. It should be noted that HST/ACS fields of giant galaxies (and, therefore, their metallicity) in some cases belong to the outskirts of the objects.
    \item The growth of the average metallicity of the red giant stars with an increase of the total absolute magnitude of the galaxies in our sample reflects the known luminosity-metallicity relation (and mass-metallicity relation). Recent studies suggest that LMR for nearby (dwarf) galaxies could have a complex nature -- mainly variations in star formation efficiency and gravitational metal retention and loss due to feedback processes like supernova and stellar winds.
    \item The majority of our sample represents dwarf galaxies which generally could be divided into dwarf spheroidals (early morphological types) and dwarf irregulars (late types). These two subsamples are clearly separated on the $M_B$--metallicity plane and the metallicity of early type galaxies is higher on average.
\end{itemize}

\begin{acknowledgments}
LNM and DGP are supported by the Russian Science Foundation grant \textnumero~24--12--00277.
\end{acknowledgments}

\bibliography{metal}{}
\bibliographystyle{aasjournal}

\end{document}